\documentclass[twocolumn,superscriptaddress,nofootinbib]{revtex4-1}
\usepackage{amsmath,amssymb,bm,graphicx,hyperref}
\allowdisplaybreaks[1]

\renewcommand\d{\partial}
\newcommand\+{\dagger}
\newcommand\<{\langle}
\renewcommand\>{\rangle}

\renewcommand\Re{\mathrm{Re}}
\renewcommand\Im{\mathrm{Im}}
\newcommand\sgn{\mathrm{sgn}}
\newcommand\G{\mathcal{G}}
\renewcommand\O{\mathcal{O}}

\newcommand\x{\bm{x}}

\begin{document}
\title{Field-theoretical aspects of one-dimensional Bose and Fermi gases\\with contact interactions}

\author{Yuta Sekino}
\affiliation{Department of Physics, Tokyo Institute of Technology, Ookayama, Meguro, Tokyo 152-8551, Japan}
\affiliation{Quantum Hadron Physics Laboratory, RIKEN Nishina Center, Wako, Saitama 351-0198, Japan}
\author{Yusuke Nishida}
\affiliation{Department of Physics, Tokyo Institute of Technology, Ookayama, Meguro, Tokyo 152-8551, Japan}

\date{\today}
\begin{abstract}
We investigate local quantum field theories for one-dimensional (1D) Bose and Fermi gases with contact interactions, which are closely connected with each other by Girardeau's Bose-Fermi mapping.
While the Lagrangian for bosons includes only a two-body interaction, a marginally relevant three-body interaction term is found to be necessary for fermions.
Because of this three-body coupling, the three-body contact characterizing a local triad correlation appears in the energy relation for fermions, which is one of the sum rules for a momentum distribution.
In addition, we apply in both systems the operator product expansion to derive large-energy and momentum asymptotics of a dynamic structure factor and a single-particle spectral density.
These behaviors are universal in the sense that they hold for any 1D scattering length at any temperature. The asymptotics for the Tonks-Girardeau gas, which is a Bose gas with a hardcore repulsion, as well as the Bose-Fermi correspondence in the presence of three-body attractions are also discussed.
\end{abstract}

\maketitle
\tableofcontents

\section{\label{sec:introduction}Introduction}
The quantum field theory (QFT) provides the description of quantum mechanics for systems with an infinite number of degrees of freedom.
This theoretical framework has been applied to different subfields of physics and revealed a variety of phenomena.
In particle physics, the standard model based on gauge principle has provided precise descriptions of three kinds of forces in nature~\cite{Weinberg:1995}.
Combined with the geometry of spacetime, QFT predicts the evaporation of black holes in cosmology~\cite{Birrell:1982}.
In condensed matter physics, QFT is used to understand excitation properties in solids~\cite{Altland:2010} as well as in ultracold atomic gases~\cite{Pethick:2008}.
The method of QFT also becomes a powerful tool to study universal physics such as critical phenomena~\cite{Zinn-Justin:1993} and low-energy excitations in spatially one-dimensional (1D) systems~\cite{Giamarchi:2004}.

Recently, QFT has been actively applied to understand universal properties of resonantly interacting systems~\cite{Braaten:2006}.
For these systems, the range $r_0$ of an interaction potential becomes much smaller than an interparticle distance, a thermal de Broglie wavelength, and a scattering length characterizing the two-body scattering at low energy.
This scale separation of $r_0$ from the other length scales leads to the universal properties of the systems, which are independent of microscopic details of the interaction.
One representative example is the universal thermodynamics of the unitary Fermi gas with an infinite scattering length~\cite{Zwerger:2012}.
Resonantly interacting systems include ultracold atoms near Feshbach resonances~\cite{Chin:2010}, dilute neutron matters~\cite{van_Wyk:2018}, and $^4\textrm{He}$ atoms~\cite{Braaten:2003}.

One of the striking features in the resonantly interacting systems is a series of exact relations called universal relations~\cite{Tan:2008,Braaten:2008a,Braaten:2012}.
These relations involve quantities called contacts, which characterize local few-body correlations, and hold for any number of particles, temperature, and scattering length as long as $r_0$ is much smaller than the other length scales.
The universal relations range from thermodynamic properties and high-energy behaviors of correlation functions such as a momentum distribution to the energy relation, which is a sum rule for the momentum distribution.
In the QFT formalism, the universal relations can be systematically derived~\cite{Braaten:2008a,Braaten:2012}.
For example, the renormalization of coupling constants leads to the energy relation.
The operator product expansion (OPE)~\cite{Kadanoff:1969,Polyakov:1970,Wilson:1969} is available to investigate correlation functions at short distance or high energy.

Resonantly interacting systems in 1D, which can be realized with ultracold atomic vapors confined into atom waveguides~\cite{Olshanii:1998,Granger:2004}, have characteristic properties.
These systems are described by models with contact interactions and they are known as integrable systems in homogeneous cases~\cite{Guan:2013,Korepin:1993}.
Another special property is a close relationship between bosons and fermions via Girardeau's Bose-Fermi mapping~\cite{Girardeau:1960}:
All the energy eigenstates of bosons interacting via an even-wave interaction with a 1D scattering length $a_B^e$~\cite{Lieb:1963} are exactly related to those of fermions interacting via an odd-wave interaction with $a_F^o=a_B^e$~\cite{Cheon:1999}.
This Bose-Fermi correspondence has been originally found in the study of the Tonks-Girardeau gas with $a_B^e\to-0$ corresponding to a noninteracting Fermi gas~\cite{Girardeau:1960}, and it has been generalized to two-component systems~\cite{Girardeau:2004}.
As a result of the Bose-Fermi correspondence, bosons and fermions with $a_B^e=a_F^o$ show the same properties in some physical quantities (see Sec.~\ref{sec:universal} for details).
On the other hand, there are of course explicit differences between these two systems.
In particular, while the even-wave interaction is well-defined without regularization, a regularization procedure is necessary to the odd-wave interaction~\cite{Cheon:1998,Cheon:1999}.
There are several ways of the regularization in the first quantized formalism~\cite{Cheon:1998,Cheon:1999,Girardeau:2004}.

The regularization of the odd-wave interaction in QFT formalism has been previously investigated to study several universal relations for fermions~\cite{Cui:2016a}.
In this QFT, fermions interact via a local two-body interaction, and the renormalization of the corresponding coupling constant is performed by solving a two-body scattering problem.
Using this renormalized coupling as well as OPE, Ref.~\cite{Cui:2016a} has derived universal relations such as power-law tails of a momentum distribution and of a radio-frequency spectroscopy as well as the adiabatic relation.
However, as shown later in this paper, a three-body problem in this QFT suffers from an ultraviolet divergence, which cannot be renormalized by the two-body coupling constant.
In addition, the energy relation derived from this theory is inconsistent with the result based on the first quantized formalism~\cite{Sekino:2018a}:
The three-body contact describing a local triad correlation is not included in the former but appears in the latter [see Eq.~(\ref{eq:energy_fermions})], while its necessity has been demonstrated in the limit of $a_F^o\to\infty$~\cite{Sekino:2018a}.
These issues imply that, besides the renormalization of the two-body coupling constant, further considerations are needed to construct QFT describing fermions with the odd-wave interaction.

In this paper, a comparative study of universal relations for 1D bosons and fermions connected with each other via the Bose-Fermi mapping is presented from the viewpoint of QFT.
In particular, we focus on analytical studies of the universal relations.
For both systems, OPE is applied to derive asymptotic behaviors of dynamic structure factors and single-particle spectral densities at large energy and momentum.
These dynamic correlation functions are important because they include information about excitations of the systems.
Also, QFT for fermions applicable to three- and higher-body problems is constructed.
We show that a marginally relevant three-body interaction term is necessary to describe fermions whose interaction is characterized only by one length scale $a_F^o$.
To demonstrate the validity of the constructed theory, we study a binding energy of three fermions and confirm that it corresponds to a three-boson bound state found by McGuire~\cite{McGuire:1964}.
In addition, we show that the three-body contact in the energy relation originates from the three-body coupling term in the QFT formalism.

This paper is structured as follows:
In Sec.~\ref{sec:universal}, we start with a brief review of 1D gases.
Section~\ref{sec:boson} is devoted to QFT for bosons to investigate dynamic correlation functions.
The quantum field theory for spinless fermions is investigated in Sec.~\ref{sec:fermion}.
We conclude this paper in Sec.~\ref{sec:conclusion}.
Our main results are universal relations for dynamic correlation functions [Eqs.~(\ref{eq:S(K)_result}), (\ref{eq:S(K)_near_TG}), (\ref{eq:quasiparticle_bosons}), (\ref{eq:Aphi(K)_result}), (\ref{eq:A_TG(K)}), (\ref{eq:Gamma_F(k)_result}), and (\ref{eq:Apsi(K)_result})] and a nonzero three-fermion coupling constant in Eq.~(\ref{eq:v3}).
Throughout this paper, the unit system of $\hbar=k_B=1$ is used and the 1D scattering lengths are set as $a_B^e=a_F^o=a$ so that the connection between bosons and fermions becomes apparent.

\section{\label{sec:universal}Bose-Fermi correspondence}
Before the discussion of QFT, we briefly review important properties of 1D Bose and Fermi gases with contact interactions in the first quantized formalism.
One way to represent contact interactions in this framework is to use pseudopotentials~\cite{Lieb:1963,Cheon:1998,Cheon:1999,Girardeau:2004}.
Interaction potentials for bosons and fermions are given by
\begin{align}\label{eq:V(x)}
V_B(x)=-\frac{2}{ma}\delta(x),\quad V_F(x)=-\frac{2a}{m}\delta'(x) D_x,
\end{align}
respectively, where $m$ is a mass of particles.
As explained above, the regularization of the contact interaction is necessary for fermions.
Here, we adopted a procedure with a regularized differential operator $D_x$~\cite{Girardeau:2004,Sekino:2018a}.
This operator acts on an $N$-body wave function as
$D_{x_{ij}}\Psi(\x)=\frac{\d}{\d x_{ij}}\Psi(\x)|_{x_{ij}=+0}$, where $\x=(x_1,\cdots,x_N)$ denotes a set of coordinates of $N$ particles and $x_{ij}=x_i-x_j$ refers to a relative coordinate between $i$th and $j$th particles.
The strengths of the pseudopotentials are characterized by the 1D scattering length $a$ and are inversely proportional to each other for bosons and fermions.

If we pick up an $N$-body wave function $\Psi_F(E;\x)$ with energy eigenvalue $E$ in the fermionic theory, there always exists its counterpart $\Psi_B(E;\x)$ with the same energy in the bosonic one.
These two wave functions are related to each other by the following Bose-Fermi mapping~\cite{Girardeau:1960}:
\begin{align}\label{eq:mapping}
\Psi_B(E;\x)=\prod_{i<j}\sgn(x_{ij})\Psi_F(E;\x),
\end{align}
where $\sgn(x)$ equals $+1$ ($-1$) for $x>0$ ($x<0$).
The inverse proportion of the interaction strengths in Eq.~(\ref{eq:V(x)}) shows that weakly (strongly) interacting bosons are mapped to strongly (weakly) interacting fermions.
Because of the correspondence in the energy spectrum, bosons and fermions with $a$, $N$, and temperature $T$ fixed share the same partition function.
As a result, all the thermodynamic quantities are identical between these two systems.
In the homogeneous cases, the ground state energy as well as the partition function at finite $T$ are well studied on the bosonic side by the method of the Bethe ansatz~\cite{Korepin:1993}.
For a repulsive interaction ($a<0$), the ground state energy and the partition function at finite $T$ in the thermodynamic limit can be exactly calculated by solving the Lieb-Liniger and Yang-Yang equations, respectively~\cite{Lieb:1963,Yang:1969}.
On the other hand, for an attractive interaction ($a>0$), there is one $N$-body bound state with energy $E=-N(N^2-1)/(6ma^2)$~\cite{McGuire:1964}.
In this paper, we focus on the thermodynamic limit for $a<0$ except for the investigation of three-fermion bound states in Sec.~\ref{sec:three-body} and Appendix~\ref{appendix:a_3}.

Since the mapping in Eq.~(\ref{eq:mapping}) never changes the absolute value of the wave functions, the two systems also share the same density correlations including static and dynamic structure factors.
In what follows, we abbreviate the label $B/F$ for physical quantities identical between bosons and fermions.
The two- and three-body contacts $C_2$ and $C_3$ can be expressed in terms of density correlations at short distances.
In order to clarify the connections of the contacts between bosons and fermions~\cite{Cui:2016a}, we here use the following definitions in which the two systems with $N$, $T$, and $a$ fixed have the same $C_2$ and $C_3$~\cite{Sekino:2018a}:
\begin{subequations}\label{eq:contacts}
\begin{align}
C_2&=\int\!dx\,\mathcal{C}_2(x)=\int\!dx\,\lim_{y\to x}\<\hat{n}(x)\hat{n}(y)\>,\\
C_3&=\int\!dx\,\mathcal{C}_3(x)=\int\!dx\,\lim_{y,z\to x}\<\hat{n}(x)\hat{n}(y)\hat{n}(z)\>,
\end{align}
\end{subequations}
where $\hat{n}(x)=\sum_{i=1}^N\delta(x-x_i)$ is the number density operator in the Schr\"odinger picture and $\<\cdots\>$ denotes a thermal average.
The two-body (three-body) contact density $\mathcal{C}_2(x)$ [$\mathcal{C}_3(x)$] measures the probability that two (three) particles come into contact with each other at the position $x$.
In the homogeneous cases, $\mathcal{C}_2(x)$ and $\mathcal{C}_3(x)$ can be exactly calculated by the Bethe ansatz~\cite{Kheruntsyan:2003,Gangardt:2003,Cheianov:2006,Kormos:2009,Kormos:2011}.

Unlike thermodynamics and density correlations, single-particle correlations such as momentum distributions and single-particle spectral densities are not forced to be identical between bosons and fermions.
Indeed, universal relations for momentum distributions reflect the difference between them.
The momentum distributions for a large momentum behave as $\rho_B(k)=4C_2/(a^2k^4)$ for bosons~\cite{Olshanii:2003} and $\rho_F(k)=4C_2/k^2$ for fermions~\cite{Cui:2016a,Sekino:2018a}.
[Here $\rho_{B/F}(k)$ are normalized as $\int (dk/(2\pi))\rho_{B/F}(k)=N$.]
Other universal relations involving $\rho_{B/F}(k)$ are energy relations.
In the absence of a trapping potential, the energy relations for bosons and fermions are given by
\begin{subequations}
\begin{align}\label{eq:energy_bosons}
E&=\int\frac{dk}{2\pi}\frac{k^2}{2m}\rho_B(k)-\frac{C_2}{ma},\\
\label{eq:energy_fermions}
E&=\int\frac{dk}{2\pi}\frac{k^2}{2m}\left(\rho_F(k)-\frac{4C_2}{k^2}\right)+\frac{C_2}{ma}+\frac{2C_3}{m},
\end{align}
\end{subequations}
respectively~\cite{Valiente:2012,Sekino:2018a}.
In the case of bosons, the interaction energy is governed by a contribution from the configuration where only two particles approach each other, leading to the last term in Eq.~(\ref{eq:energy_bosons}).
On the other hand, the effect of the configuration where a trio of particles approach each other is not negligible for fermions and thus $C_3$ appears in the energy relation.
This situation is similar to the 3D cases with the Efimov effect~\cite{Efimov:1970} in the sense that three-body correlations cannot be neglected in the energy relation~\cite{Castin:2011,Braaten:2011}.
As mentioned in Sec.~\ref{sec:introduction}, the necessity of $C_3$ in Eq.~(\ref{eq:energy_fermions}) was demonstrated in the limit of $a\to\infty$~\cite{Sekino:2018a}.

In the next two sections to discuss QFT, we use the following shorthand notations:
The differential $\tensor{\partial}$ is defined by $A\tensor{\partial}B\equiv [A\d B-(\d A)B]/2,$ $X=(t,x)$ refers to a spacetime coordinate, and $K=(K_0,K_1)=(\omega,k)$ to a set of energy $\omega$ and momentum $k$.
The inner product between $K$ and $X$ is given by $K\cdot X=\omega t-kx$, and $AB\cdots C(X)\equiv A(X)B(X)\cdots C(X)$ is assumed.

\section{\label{sec:boson}Bosons}
This section is devoted to studies of high-energy behaviors of dynamic correlation functions for 1D bosons.
The Lagrangian density for bosons with an even-wave interaction is given by
\begin{align}\label{eq:L_B}
\mathcal{L}_B=\phi^\+\left(i\d_t+\frac{\d_x^2}{2m}\right)\phi+\frac{1}{ma}\phi^\+\phi^\+\phi\phi,
\end{align}
where $\phi$ is a bosonic field.
For convenience of diagrammatic calculations, we perform the Hubbard-Stratonovich transformation~\cite{Altland:2010}.
The transformed Lagrangian density is
\begin{align}\label{eq:L'_B}
\mathcal{L}'_B=\phi^\+\left(i\d_t+\frac{\d_x^2}{2m}\right)\phi-\frac{1}{ma}(\Phi^\+\Phi-\Phi^\+\phi^2-\phi^{\+2}\Phi),
\end{align}
where $\Phi$ is introduced as an auxiliary bosonic field.
Because the Euler-Lagrange equation for $\Phi^\+$ provides $\Phi=\phi^2$, $\Phi$ has the degree of freedom of a dimer.
In terms of field operators, the number density operator in the Heisenberg picture is written as $\hat{n}(X)=\phi^\+\phi(X)$. 
In this bosonic theory, there is no renormalization of composite operators, and thus we can naively take the limit in Eqs.~(\ref{eq:contacts}):
\begin{subequations}
\begin{align}
\mathcal{C}_2&=\<\phi^{\+2}\phi^2(X)\>=\<\Phi^\+\Phi(X)\>,\\
\mathcal{C}_3&=\<\phi^{\+3}\phi^3(X)\>=\<\Phi^\+\phi^\+\phi\Phi(X)\>.
\end{align}
\end{subequations}
Note that the contact densities of a thermal equilibrium state are independent of $X$ because the system is translationally invariant in spacetime.

\begin{figure}[t]
\includegraphics[width=\columnwidth,clip]{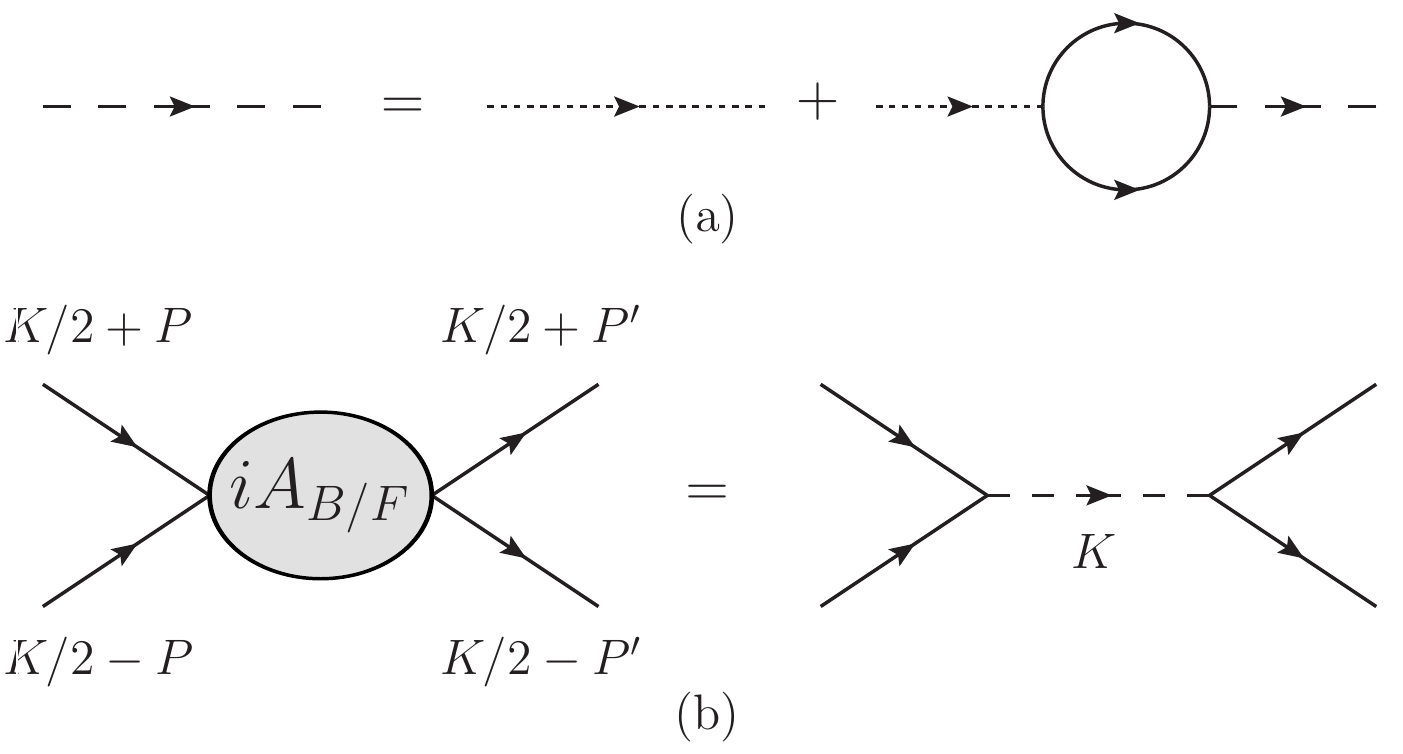}
\caption{\label{fig:A&D}
Feynman diagrams for (a) the full dimer propagator and (b) the two-body scattering amplitude.
The solid, dotted, and dashed lines indicate $iG(K)$, $iD_{B/F}^{(0)}$, and $iD_{B/F}(K)$, respectively.}
\end{figure}
We now present notations in Feynman diagrams for later diagrammatic calculations.
The propagator $iG(K)=iG(\omega,k)$ of a boson with energy $\omega$ and momentum $k$ is denoted by a solid line and given by
\begin{align}\label{eq:G(K)}
G(K)=\frac{1}{\omega-\frac{k^2}{2m}+i0^+}.
\end{align}
On the other hand, a dashed (dotted) line denotes a full (bare) propagator $iD_B(K)$ ($iD_B^{(0)}=-ima$) of a dimer.
Solving the Dyson equation in Fig.~\ref{fig:A&D}(a), where the boson-dimer vertex is $2i/(ma)$, we obtain
\begin{align}\label{eq:D_B(K)}
D_B(K)=-\frac{ma}{1-1/(a\beta_K)}
\end{align}
with $\beta_K\equiv\sqrt{k^2/4-m\omega-i0^+}$.
Unlike contact interactions in higher dimensions, $D_B(K)$ is obtained without regularization.
The scattering amplitude of two bosons $iA_B(K)$ depicted in Fig.~\ref{fig:A&D}(b) is related to $D_B(K)$ through
\begin{align}\label{eq:A_B(K)}
A_B(K)=-\frac{4}{m^2a^2}D_B(K)=\frac{4}{m}\frac{1}{a-1/\beta_K}.
\end{align}
Note that incoming and outgoing bosons have the same total energy $\omega$ and center-of-mass momentum $k$ because of the energy and momentum conservations.

The dynamic structure factor $S(K)$ and the single-particle spectral density $\mathcal{A}_B(K)$ are defined as the imaginary parts of retarded response functions:
\begin{align}\label{eq:S(K)}
S(K)&=-\frac{1}{\pi}\frac{\Im\left[\<\G^{\mathrm{R}}_{\hat{n}}(K)\>\right]}{1-e^{-\omega/T}},\\
\label{eq:Aphi(K)}
\mathcal{A}_B(K)&=-\frac{1}{\pi}\Im\left[\<\G^{\mathrm{R}}_\phi(K)\>\right],
\end{align}
where $\G^{\mathrm{R}}_A(K)=-i\int\!d^2X\,e^{iK\cdot X-\,0^+t}\theta(t)[A(X),A^\+(0)]$ and $\theta(t)$ is the Heaviside step function.
In the QFT framework, it is more convenient to calculate time-ordered Green's functions defined by
\begin{align}\label{eq:<GA>2}
\<\G_A(K)\>=-i\int\!d^2X\,e^{iK\cdot X}\<\mathcal{T}[A(X)A^\+(0)]\>
\end{align}
than $\<\G^{\mathrm{R}}_A(K)\>$ because diagrammatic calculations are directly applicable.
From the Lehmann representations, Eqs.~(\ref{eq:S(K)}) and (\ref{eq:Aphi(K)}) are rewritten as
\begin{align}\label{eq:S(K)2}
S(K)&=-\frac{1}{\pi}\Im\left[\<\G_{\hat{n}}(K)\>\right]+O(e^{-\omega/T}),\\
\label{eq:Aphi(K)2}
\mathcal{A}_B(K)&=-\frac{1}{\pi}\Im\left[\<\G_\phi(K)\>\right]+O(e^{-\omega/T}).
\end{align}
We will study $S(K)$ and $\mathcal{A}_B(K)$ at high energy $\omega>0$ and large momentum $|k|$ by using OPE.

In this section, we proceed as follows:
We begin with the introduction of OPE in Sec.~\ref{sec:OPE}.
Then OPE is applied to the density (single-particle) Green's function in Sec.~\ref{sec:density} (Sec.~\ref{sec:particle}).
The asymptotic behaviors of the dynamic structure factor and the single-particle spectral density at large energy and momentum are discussed in Sec.~\ref{sec:S(K)} and Sec.~\ref{sec:A_B(K)}, respectively.

\subsection{\label{sec:OPE}Operator product expansion}
In QFT, OPE states that the product of two operators $A(X)$ and $B(0)$ at different spacetime points can be given by a sum of local operators $\O$ at $X=0$~\cite{Kadanoff:1969,Polyakov:1970,Wilson:1969}:
\begin{align}\label{eq:A(X)B(0)}
A(X)B(0)=\sum_\O w^\O(X)\O(0).
\end{align}
Hereafter, a shorthand notation $\O=\O(X=0)$ is used.
The quantities $w^\O(X)$ called Wilson coefficients are c-number functions of $X$.
Such an operator product appears in studies of static or dynamic correlation functions including $S(K)$ and $\mathcal{A}_B(K)$.
From OPE in Eq.~(\ref{eq:A(X)B(0)}), $\G_A(K)$ in Eq.~(\ref{eq:<GA>2}) can be expressed as
\begin{align}\label{eq:OPE}
\G_A(K)=\sum_{\O}W_{A}^{\O}(K)\O.
\end{align}
The dependences of Wilson coefficients on the operator $A$ are explicitly shown for later convenience.

The operator product expansion becomes a powerful tool to study $\<\G_A(K)\>$ at large energy and momentum, i.e., in a region where $\sqrt{m|\omega|}$ and $|k|$ are much larger than typical scales of a given state such as $n=\<\hat{n}\>$ and $\sqrt{mT}$.
To see the usefulness of OPE, let us take thermal averages of both sides of Eq.~(\ref{eq:OPE}).
By dimensional analysis, $\<\G_A(K)\>$ is expressed as
\begin{align}\label{eq:<G_A>}
\<\G_A(K)\>=\sum_{\O}\frac{1}{k^{\Delta_{\O}+3-2\Delta_A}}f_A^{\O}\left(\frac{k^2}{2m\omega}\right)\<\O\>,
\end{align}
where $f_A^{\O}(k^2/(2m\omega))$ is a dimensionless function. 
The scaling dimension $\Delta_\O$ is defined so that the equal-time correlation function $\<\O(0,x)\O^\+(0)\>$ with small separation $x$ behaves as $1/|x|^{2\Delta_\O}$.
In our counting scheme, dimensions of particle mass, momentum, and energy are counted as $0$, $1$, and $2$, respectively.
Equation~(\ref{eq:<G_A>}) shows that Wilson coefficients with small $\Delta_\O$ dominate $\<\G_A(K)\>$ at large energy and momentum.

Since OPE is an operator identity, the expansion of $\<\G_A(K)\>$ for large $K$ [Eq.~(\ref{eq:<G_A>})] is valid for any average $\<\cdots\>$, i.e., universal in the sense that it is independent of details of a given many-body state such as a number density and a temperature.
In nonrelativistic QFT, $W_{A}^{\O}(K)$ of local operators with small $\Delta_\O$ can be determined by solving few-body problems as shown below.
On the other hand, information specific to the given many-body state is encoded in local physical quantities $\<\O\>$.

In the case of 1D bosons described by the Lagrangian density~(\ref{eq:L'_B}), the dimensionless coefficients $f_A^{\O}(k^2/(2m\omega))$ in Eq.~(\ref{eq:<G_A>}) depend not only on $k^2/(2m\omega)$ but also on a scaled interaction strength $1/(ka)$.
Therefore, the behavior of $f_A^{\O}(k^2/(2m\omega),1/(ka))$ for large $|k|$ with $k^2/(2m\omega)$ and $a$ fixed is equivalent to that for large $|a|$ with $k^2/(2m\omega)$ and $k$ fixed except for the case of a vanishing scattering length.
Recalling Eq.~(\ref{eq:V(x)}) or (\ref{eq:L_B}), we see that the limit of an infinite scattering length corresponds to the noninteracting limit.
A perturbative few-body calculation is thus available to derive the large-$K$ behavior of $\<\G_A(K)\>$.
Since the power-law tail of $\<\G_A(K)\>$ results from the interaction, the dimensionless coefficients for small $1/(ka)$ with $k^2/(2m\omega)$ fixed can be expanded as
\begin{align}
f_A^{\O}\left(\frac{k^2}{2m\omega},\frac{1}{ka}\right)=\left(\frac1{ka}\right)^{N_A^\O}g_A^{\O}\left(\frac{k^2}{2m\omega}\right)+\cdots
\end{align}
with $N_A^\O>0$.
This means that the power-law decays of $W_{A}^{\O}(K)$ are shifted by $N_A^\O$ from the estimation in Eq.~(\ref{eq:<G_A>}) based on scaling dimensions.
This situation in 1D bosons with a finite interaction strength is similar to that in 1D two-component fermions~\cite{Barth:2011}.
We note that the perturbative few-body analysis of the Wilson coefficients does not mean the perturbative treatment of the many-body state because the expectation values $\<\O\>$ in Eq.~(\ref{eq:<G_A>}) depend nonperturbatively on a dimensionless coupling constant $\gamma=-2/(na)$, which characterizes the interaction strength of 1D bosons with the number density $n$.
On the other hand, the above analysis cannot be applied to 1D fermions with an odd-wave interaction studied in Sec.~\ref{sec:fermion}.
In this case, the limit of an infinite scattering length is no longer the weakly interacting limit as in the 3D cases with $s$-wave interactions.
Thus, $N_A^\O>0$ is not imposed and some nonperturbative treatment is necessary to compute $W_{A}^{\O}(K)$.
Similarly, such a shift is not imposed on the Tonks-Girardeau gas with a hardcore repulsion ($-1/a\to+\infty$).

In the following subsections, we apply OPE to the derivations of large-$K$ tails of density and single-particle Green's functions.
Since the two-body contact density $\mathcal{C}_2=\<\Phi^\+\Phi\>$ is a central quantity in the context of universal relations, we focus on how $\mathcal{C}_2$ affects these correlation functions at large energy and momentum.
In order to determine the Wilson coefficient of $\Phi^\+\Phi$, we have to take the following local operators into account:
the unit operator
\begin{align}\label{eq:unit_operator}
1
\end{align}
with $\Delta_1=0$, one-body operators
\begin{align}\label{eq:Obc}
\O_{b,c}=\phi^\+(i\tensor{\partial}_t)^b(-i\tensor{\partial}_x)^c\phi
\end{align}
with $\Delta_{\O_{b,c}}=2b+c+1\leq4$, and a dimer density operator
\begin{align}\label{eq:d^+d}
\Phi^\+\Phi\sim \phi^\+\phi^\+\phi\phi
\end{align}
with $\Delta_{\Phi^\+\Phi}=2.$
We note that local operators with total derivatives are not considered because their thermal averages vanish by the translational invariance of the system.
Similarly, thermal averages of $\O_{b,c}$ with odd $c$ also vanish due to the inversion invariance, while these $\O_{b,c}$ should be taken into account in few-body calculations to determine the Wilson coefficient of $\Phi^\+\Phi$.

The Wilson coefficients of above local operators can be determined by the following matching procedure:
First, matrix elements of these local operators $\O$ with respect to states $\<\mu|$ and $|\nu\>$ are computed.
Second, the matrix elements $\<\mu|\G_A(K)|\nu\>$ are calculated and expanded in small momentum scales $P_\mathrm{ex}$ associated with the external states.
Then, by demanding that the expansions of both sides of Eq.~(\ref{eq:OPE}) match in each order of $P_\mathrm{ex}$, the coefficients $W_{A}^{\O}(K)$ are determined.
Because OPE is an operator identity, the simplest states for which $\<\mu|\O|\nu\>$ is nonzero can be used to determine $W_A^{\O}(K)$.
For instance, the vacuum state $|\mathrm{vac}\>$ with no particle is available to determine the coefficient of the unit operator.
Because of $\<\mathrm{vac}|\O|\mathrm{vac}\>=0$ for $\O\neq1$ on the right-hand side of Eq.~(\ref{eq:OPE}), taking vacuum expectation values of both sides yields $W_A^1(K)=\<\mathrm{vac}|\G_A(K)|\mathrm{vac}\>$.
Similarly, expectation values with respect to a one-boson (two-boson) state are used to compute the coefficients for one-body operators $\O_{b,c}$ (the dimer density operator $\Phi^\+\Phi$).

\subsection{\label{sec:density}OPE for $\G_{\hat{n}}(K)$}
We now apply OPE to the dynamic structure factor $S(K)$.
By substituting
\begin{align}\label{eq:Gn_OPE}
\G_{\hat{n}}(K)=\sum_{\O}W_{\hat{n}}^{\O}(K)\O
\end{align}
into Eq.~(\ref{eq:S(K)2}), $S(K)$ can be expanded as
\begin{align}\label{eq:S(K)_OPE}
S(K)=-\frac{1}{\pi}\sum_{\O}\textrm{Im}\left[W_{\hat{n}}^{\O}(K)\right]\<\O\>.
\end{align}
We note that all the local operators $\O$ which we take into account are Hermitian [see Eqs.~(\ref{eq:unit_operator})--(\ref{eq:d^+d})], leading to real-valued $\<\O\>$.

\begin{figure}[t]
\includegraphics[width=\columnwidth,clip]{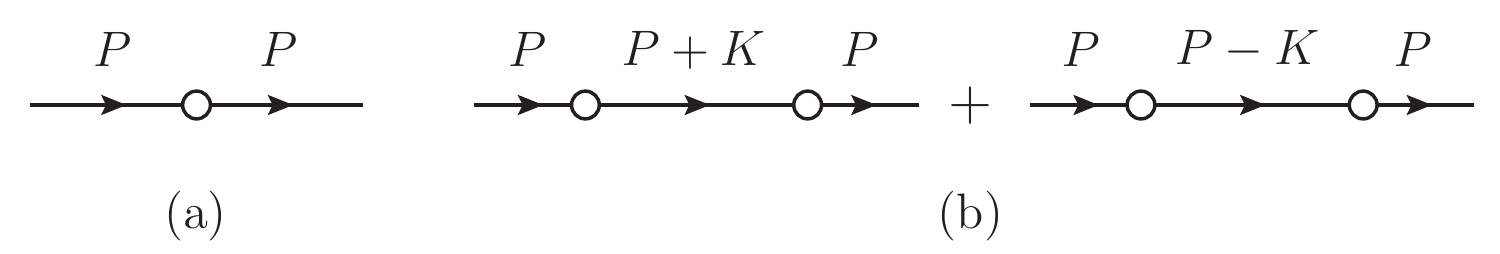}
\caption{\label{fig:<1|O|1>_<1|Gnn|1>}
Diagrams for the expectation values of (a) $\O_{b,c}$ and of (b) $\G_{\hat{n}}(K)$ with respect to a one-boson state $|\phi_P\>$.
The open dot in (a) denotes the insertion of $\O_{b,c}$, while those in (b) denote the insertions of the density operators in $\G_{\hat{n}}(K)$.}
\end{figure}

In order to determine $W_{\hat{n}}^\O(K)$ for operators in Eqs.~(\ref{eq:unit_operator})--(\ref{eq:d^+d}), we employ the matching procedure explained above.
Taking the vacuum expectation values of both sides of Eq.~(\ref{eq:Gn_OPE}), we find $W_{\hat{n}}^1(K)=\<\mathrm{vac}|\G_{\hat{n}}(K)|\mathrm{vac}\>=0$.
The coefficients of $\O_{b,c}$ are derived by evaluating both sides of Eq.~(\ref{eq:Gn_OPE}) with respect to a one-boson state $|\phi_P\>=|\phi_{(P_0,P_1)}\>$, in which the boson has energy $P_0$ and momentum $P_1$.
Here, we do not impose the on-shell condition, i.e., $P_0\neq P_1^2/(2m)$.
The expectation values of $\O_{b,c}$ on the right-hand side equal
\begin{align}\label{eq:<1|One|1>_boson}
\<\phi_P|\O_{b,c}|\phi_P\>=(P_0)^b(P_1)^c,
\end{align}
which can be expressed in terms of the Feynman diagram as Fig.~\ref{fig:<1|O|1>_<1|Gnn|1>}(a).
On the other hand, the expectation value of $\G_{\hat{n}}(K)$ on the left-hand side is given by the diagrams in Fig.~\ref{fig:<1|O|1>_<1|Gnn|1>}(b) and equals
\begin{align}\label{eq:<1|G_nn|1>}
\<\phi_P|\G_{\hat{n}}(K)|\phi_P\>=G(P+K)+G(P-K).
\end{align}
By comparing its expansion in $P$ with Eq.~(\ref{eq:<1|One|1>_boson}), the coefficients are determined by
\begin{align}\label{eq:WnnOab_in-general}
W_{\hat{n}}^{\O_{b,c}}(K)
=\frac{1}{b!c!}\frac{\partial^{b+c}G(K)}{\partial \omega^b\partial k^c}+(K\to-K).
\end{align}
Because of $\Im[G(K)]=-\pi\delta(\omega-k^2/(2m))$, all the $W_{\hat{n}}^{\O_{b,c}}(K)$ in Eq.~(\ref{eq:S(K)_OPE}) contribute to $S(K)$ for $\omega>0$ only at the single-particle peak $\omega=k^2/(2m)$.

\begin{figure}[t]
\includegraphics[width=\columnwidth,clip]{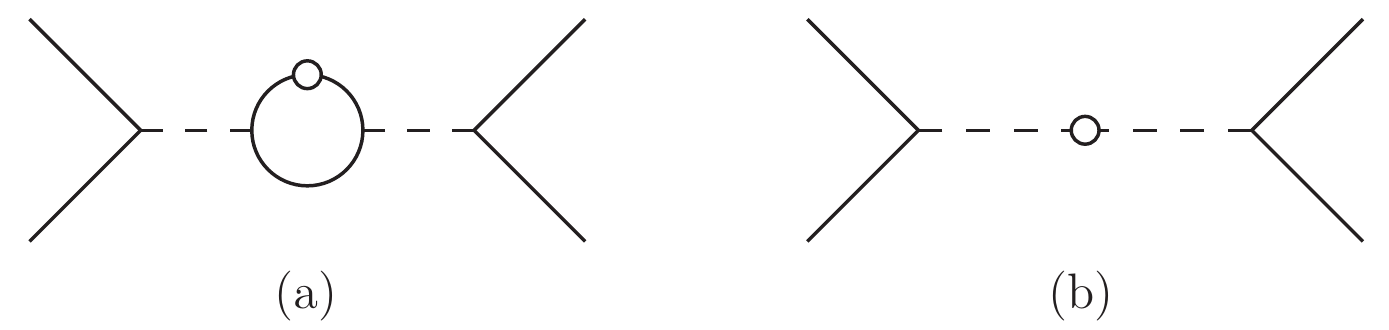}
\caption{\label{fig:<2|O|2>}
Diagrams for the expectation values of (a) $\mathcal{O}_{b,c}$ and of (b) $\Phi^\+\Phi$ with respect to a two-boson state $|\phi_{P/2}^2\>$.}
\end{figure}
To determine the coefficient of $\Phi^\+\Phi$, we next calculate the expectation values of both sides of Eq.~(\ref{eq:Gn_OPE}) with respect to an off-shell two-boson state $|\phi_{P/2}^2\>$, in which the two bosons have the same energy and momentum.
For a shorthand notation, we define $\<\cdots\>_2\equiv\<\phi_{P/2}^2|\cdots|\phi_{P/2}^2\>$.
With the help of Eq.~(\ref{eq:WnnOab_in-general}) determined in the one-boson sector, several terms from $\<\O_{b,c}\>_2$ on the right-hand side automatically match terms in $\<\G_{\hat{n}}(K)\>_2$ on the left-hand side.
These terms correspond to diagrams where all the operators are inserted into one external line and do not affect the determination of $W_{\hat{n}}^{\Phi^\+\Phi}(K)$.
In addition, diagrams for $\<\G_{\hat{n}}(K)\>_2$ in which the two density operators are inserted into different external lines without interactions contribute to $W_{\hat{n}}^{\Phi^\+\Phi}(K)$ only at $K=0$.
For these reasons, we below consider the other diagrams which are necessary to determine $W_{\hat{n}}^{\Phi^\+\Phi}(K)$ for nonzero $K$.

Let us now calculate the expectation values of local operators on the right-hand side of Eq.~(\ref{eq:Gn_OPE}).
Figures~\ref{fig:<2|O|2>}(a) and \ref{fig:<2|O|2>}(b) show diagrams contributing to the expectation values of $\O_{b,c}$ with $\Delta_{\O_{b,c}}\leq4$ and $\Phi^\+\Phi$, respectively, and thus we obtain
\begin{align}
\<\O_{b,c}\>_2&=[A_B(P)]^2I_{b,c}(P),\\
\<\Phi^\+\Phi\>_2&=\frac{m^2a^2}{4}[A_B(P)]^2,
\end{align}
where integrals corresponding to the loop in Fig.~\ref{fig:<2|O|2>}(b) are given by
\begin{align}\label{eq:IOab}
I_{b,c}(P)=i\int_QG(Q)[G(P-Q)]^2(P_0-Q_0)^b(P_1-Q_1)^c
\end{align}
with $\int_Q\equiv\int dQ_0dQ_1/(2\pi)^2$.
These integrals can be analytically computed and their explicit forms are shown in Appendix~\ref{appendix:integrals} [see Eqs.~(\ref{eq:I_O_bc})].
Finally, the expectation value of the right-hand side of Eq.~(\ref{eq:Gn_OPE}) divided by $[A_B(P)]^2$ is found to be
\begin{align}\label{eq:<2|Gnn|2>_RHS}
&\sum_{\O}\frac{W_{\hat{n}}^{\O}(K)\<\O\>_2}{[A_B(P)]^2}\nonumber\\
&=\sum_{\Delta_{\O_{b,c}}\leq4}W_{\hat{n}}^{\O_{b,c}}(K)I_{b,c}(P)
+\frac{m^2a^2}{4}W_{\hat{n}}^{\Phi^\+\Phi}(K)+O(P)
\end{align}
with $O(P)\equiv O(\beta_P)+O(P_1)$.
Higher-order contributions come from higher derivative local operators and they vanish in the limit of $P\to0$.

\begin{figure}[t]
\includegraphics[width=\columnwidth,clip]{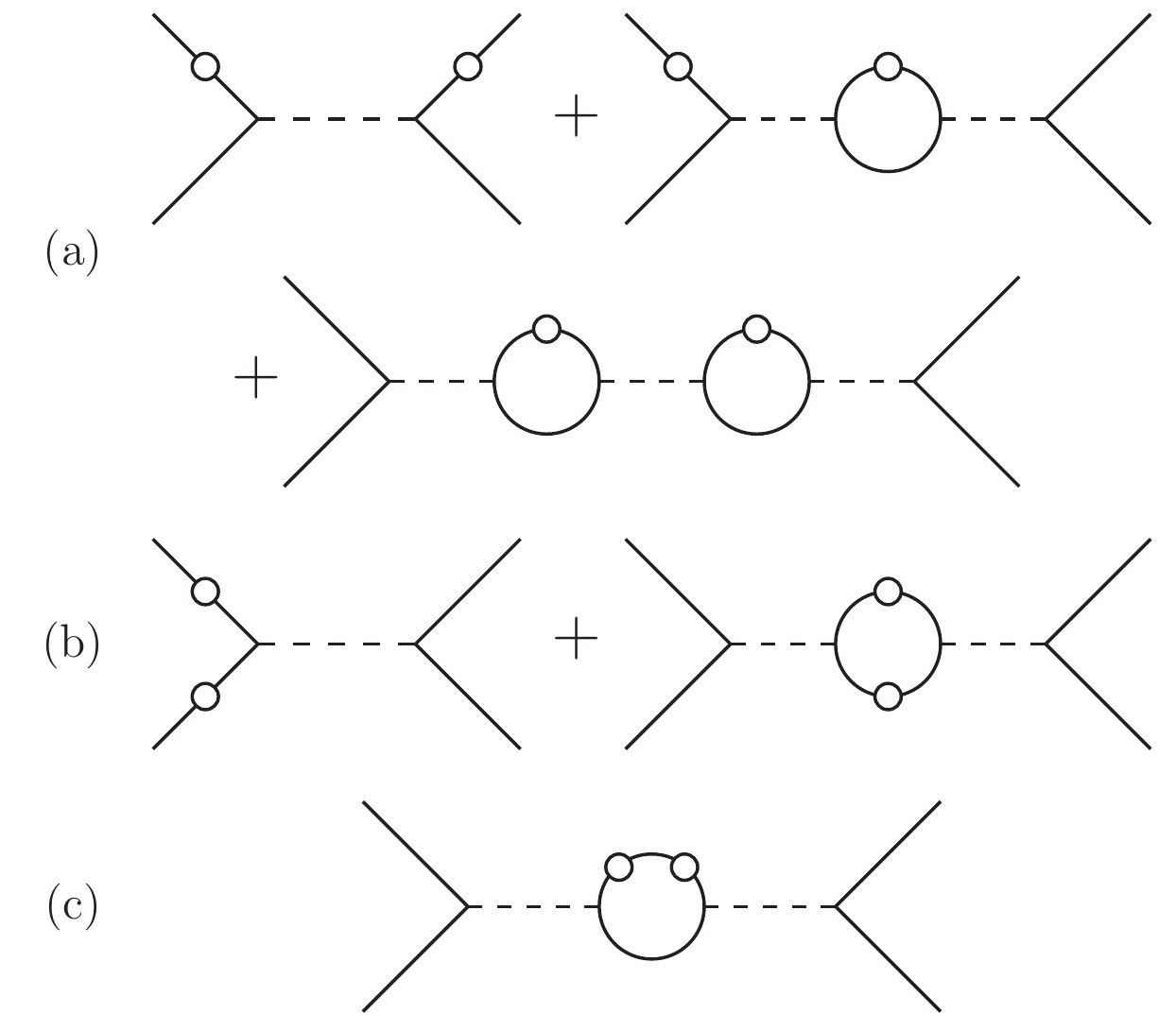}
\caption{\label{fig:<2|G_n|2>}
Graph topologies contributing to the expectation value $\<\G_{\hat{n}}(K)\>_2$.}
\end{figure}
We turn to the expectation value $\<\G_{\hat{n}}(K)\>_2$ on the left-hand side of Eq.~(\ref{eq:Gn_OPE}).
Diagrams contributing to $\<\G_{\hat{n}}(K)\>_2$ are depicted in Fig.~\ref{fig:<2|G_n|2>}.
These contributions are given by
\begin{subequations}\label{eq:<Gn>(a)--(c)}
\begin{align}\label{eq:<Gn>(a)}
\frac{\<\G_{\hat{n}}(K)\>_\textrm{(a)}}{[A_B(P)]^2}
&=-A_B(K+P)\left(\frac{2G(K+P/2)}{A_B(P)}-J_1(K,P)\right)^2\nonumber\\
&\quad+(K\to-K),\\
\label{eq:<Gn>(b)}
\frac{\<\G_{\hat{n}}(K)\>_\textrm{(b)}}{[A_B(P)]^2}
&=-\frac{4G(K+P/2)G(-K+P/2)}{A_B(P)}-J_2(K,P),\\
\label{eq:<Gn>(c)}
\frac{\<\G_{\hat{n}}(K)\>_\textrm{(c)}}{[A_B(P)]^2}&=J_3(K,P)+(K\to-K),
\end{align}
\end{subequations}
where integrals corresponding to loops in Fig.~\ref{fig:<2|G_n|2>} are given by
\begin{subequations}\label{eq:Js_def}
\begin{align}
J_1(K,P)&= i\int_QG(Q)G(P-Q)G(K+P-Q),\\
J_2(K,P)&=-i\int_QG(Q)G(Q+K)G(P-Q)\nonumber\\
&\quad\times G(P-K-Q),\\
J_3(K,P)&= i\int_Q G(Q)[G(P-Q)]^2G(P+K-Q).
\end{align}
\end{subequations}
The analytical expressions of these integrals are shown in Appendix~\ref{appendix:integrals} [see Eqs.~(\ref{eq:Js})].
As shown in Eqs.~(\ref{eq:Js_exp}), Eqs.~(\ref{eq:Js_def}) can be expanded in $P$ as
\begin{subequations}\label{eq:Js_exp_main}
\begin{align}
J_1(K,P)&=-\frac{m}{2\beta _P}\left(G(K)+\frac{kP_1[G(K)]^2}{2m}\right)\nonumber\\
&\quad-\beta_K[G(K)]^2-\frac{mG(K)}{2\beta _{K}}+O(P),\\
J_2(K,P)
&=\frac{m}{\beta_P}\left.\left.\left(G(K)+\frac{kP_1[G(K)]^2}{2m}\right)\right(K\to-K\right)\nonumber\\
&\quad+\frac{m[G(K)]^2}{2\beta_{K}}+\frac{m[G(-K)]^2}{2\beta_{-K}}+O(P),
\\
J_3(K,P)&=\frac{[G(K)]^4}{2m}\left(k^2\beta_K-\frac{(m\omega)^2}{\beta_K}\right)\nonumber\\
&\quad+\sum_{\Delta_{\O_{b,c}}\leq4}\frac{1}{b!c!}\frac{\d^{b+c}G(K)}{\d\omega^b\d k^c}I_{b,c}(P)+O(P).
\end{align}
\end{subequations}
The expansion of $\<\G_{\hat{n}}(K)\>_2/[A(P)]^2$ in $P$ can be performed by summing up Eqs.~(\ref{eq:<Gn>(a)--(c)}) and substituting Eqs.~(\ref{eq:Js_exp_main}) into the sum.
The result is
\begin{align}\label{eq:<2|Gnn|2>_LHS_expanded}
\frac{\<\G_{\hat{n}}(K)\>_2}{[A_B(P)]^2}
&=\frac{m^3a}{4}\left[\frac{1}{1-a\beta_K}\left(\frac{kG(K)}{m}\right)^4+\frac{4}{k^2}\frac{G(K)}{m}\right.\nonumber\\
&\qquad\qquad\left.-4\left(\frac{G(K)}{m}\right)^2+(K\to-K)\right]\nonumber\\
&\quad+\sum_{\Delta_{\O_{b,c}}\leq4}W_{\hat{n}}^{\O_{b,c}}(K)I_{b,c}(P)+O(P).
\end{align}
By comparing this with Eq.~(\ref{eq:<2|Gnn|2>_RHS}) in the limit of $P\to0$, the Wilson coefficient of $\Phi^\+\Phi$ is found to be
\begin{align}\label{eq:Wnnd^+d(K)}
W_{\hat{n}}^{\Phi^\+\Phi}(K)
&=\frac{m}{a}\left[\frac{1}{1-a\beta_K}\left(\frac{kG(K)}{m}\right)^4+\frac{4}{k^2}\frac{G(K)}{m}\right.\nonumber\\
&\qquad\quad\left.-4\left(\frac{G(K)}{m}\right)^2+(K\to-K)\right].
\end{align}

\subsection{\label{sec:S(K)}Dynamic structure factor}
We now evaluate the large-energy and momentum behavior of $S(K)$ [Eq.~(\ref{eq:S(K)_OPE})] away from the single-particle peak.
As shown in the previous subsection, there is no contribution from the one-body operators to $S(K)$ for $\omega\neq k^2/(2m)$.
The imaginary part of $W_{\hat{n}}^{\Phi^\+\Phi}(K)$ thus dominates $S(K)$ in the large-$K$ limit:
\begin{align}\label{eq:S(K)_OPE_2}
S(K)=-\frac{1}{\pi}\textrm{Im}\left[W_{\hat{n}}^{\Phi^\+\Phi}(K)\right]\mathcal{C}_2+O\left(K^{-7}\right)
\end{align}
with $\mathcal{C}_2=\<\Phi^\+\Phi\>$.
The corrections come from two-body operators with derivatives as well as higher-body operators and their orders can be estimated with the help of the perturbation theory.
From Eq.~(\ref{eq:Wnnd^+d(K)}), the imaginary part of $W_{\hat{n}}^{\Phi^\+\Phi}(K)$ reads
\begin{align}\label{eq:Im[Wnnd^+d(K)]}
&-\frac{1}{\pi}\Im\left[W_{\hat{n}}^{\Phi^\+\Phi}(K)\right]\nonumber\\
&=\theta(m\omega-k^2/4)\frac{\left[kG(K)\right]^4}{\pi m^3}
\frac{\sqrt{m\omega-k^2/4}}{1+a^2(m\omega-k^2/4)}.
\end{align}
The Heaviside step function $\theta(m\omega-k^2/4)$ represents the two-particle threshold, which is also pointed out in the 2D and 3D cases~\cite{Son:2010,Hofmann:2011}.
This threshold reflects the fact that the excitations of two particles with center-of-mass momentum $k$ require energies larger than their center-of-mass energy $k^2/(4m)$.

Substituting the expansion of Eq.~(\ref{eq:Im[Wnnd^+d(K)]}) in large $K$ into Eq.~(\ref{eq:S(K)_OPE_2}), we can obtain the following power-law behavior of $S(K)$ above the two-particle threshold:
\begin{align}\label{eq:S(K)_result}
S(\omega,k)=\frac{m}{\pi a^2}\left(\frac{k}{m\omega-k^2/2}\right)^4\frac{\mathcal{C}_2}{\sqrt{m\omega-k^2/4}}.
\end{align}
This behavior holds when $\sqrt{m\omega}$ and $|k|$ are much larger than $n=\<\hat{n}\>$, $1/|a|$, and $\sqrt{mT}$.
As mentioned earlier, $\mathcal{C}_2$ can be exactly calculated for any scattering length and temperature by the Bethe ansatz.
By combining this exact result of $\mathcal{C}_2$ with Eq.~(\ref{eq:S(K)_result}), $S(K)$ at large $K$ can be completely determined for any scattering length ($0<-1/a<+\infty$) and temperature ($T\geq0$).
From the Bose-Fermi correspondence, this power law in $S(K)$ holds for 1D fermions with an odd-wave interaction.
We note that our result [Eq.~(\ref{eq:S(K)_result})] is not valid for the Tonks-Girardeau gas with a hardcore repulsion ($-1/a\to+\infty$) because the expansion of $W_{\hat{n}}^{\Phi^\+\Phi}(K)$ in small $1/|a|\ll\sqrt{m\omega},|k|$ was used.
Indeed, the Bose-Fermi correspondence makes $S(K)$ of the Tonks-Girardeau gas identical to that of free fermions, which has no power-law tail at large $K$.
Similarly, the tail of $S(K)$ vanishes in the noninteracting limit ($1/a\to0$).

Near the Tonks-Girardeau limit ($0<n|a|\ll1$), $S(K)$ also shows another power-law behavior above the two-particle threshold.
Substituting the expansion of Eq.~(\ref{eq:Im[Wnnd^+d(K)]}) in small $a$ into Eq.~(\ref{eq:S(K)_OPE_2}), we obtain
\begin{align}\label{eq:S(K)_near_TG}
S(\omega,k)=\frac{m\sqrt{m\omega-k^2/4}}{\pi}\left(\frac{k}{m\omega-k^2/2}\right)^4\mathcal{C}_2.
\end{align}
This behavior holds when $\sqrt{m\omega}$ and $|k|$ are much larger than $n$ and $\sqrt{mT}$ but much smaller than $1/|a|$.
For $m\omega\gg k^2$, the asymptotic behavior in Eq.~(\ref{eq:S(K)_near_TG}) reduces to $S(\omega,k)\sim k^4/\omega^{7/2}$, which is consistent with the recent result based on the Bethe ansatz~\cite{Granet:2020}.
We note that, when $\sqrt{m\omega}$ and $|k|$ become much larger than $1/|a|$, $S(K)$ should again obey Eq.~(\ref{eq:S(K)_result}) even near the Tonks-Girardeau limit.

\subsection{\label{sec:particle}OPE for $\G_\phi(K)$}
Next, we will apply OPE of field operators,
\begin{align}\label{eq:Gphi_OPE}
\G_\phi(K)=\sum_{\O}W_\phi^\O(K)\O,
\end{align}
to study the large-$K$ behavior of the single-particle spectral density $\mathcal{A}_B(K)$ in Eq.~(\ref{eq:Aphi(K)2}).
The coefficients $W_\phi^\O(K)$ for the local operators in Eqs.~(\ref{eq:unit_operator})--(\ref{eq:d^+d}) can be determined by the matching procedure in a similar way as in Sec.~\ref{sec:density}.

\begin{figure}[t]
\includegraphics[width=0.9\columnwidth,clip]{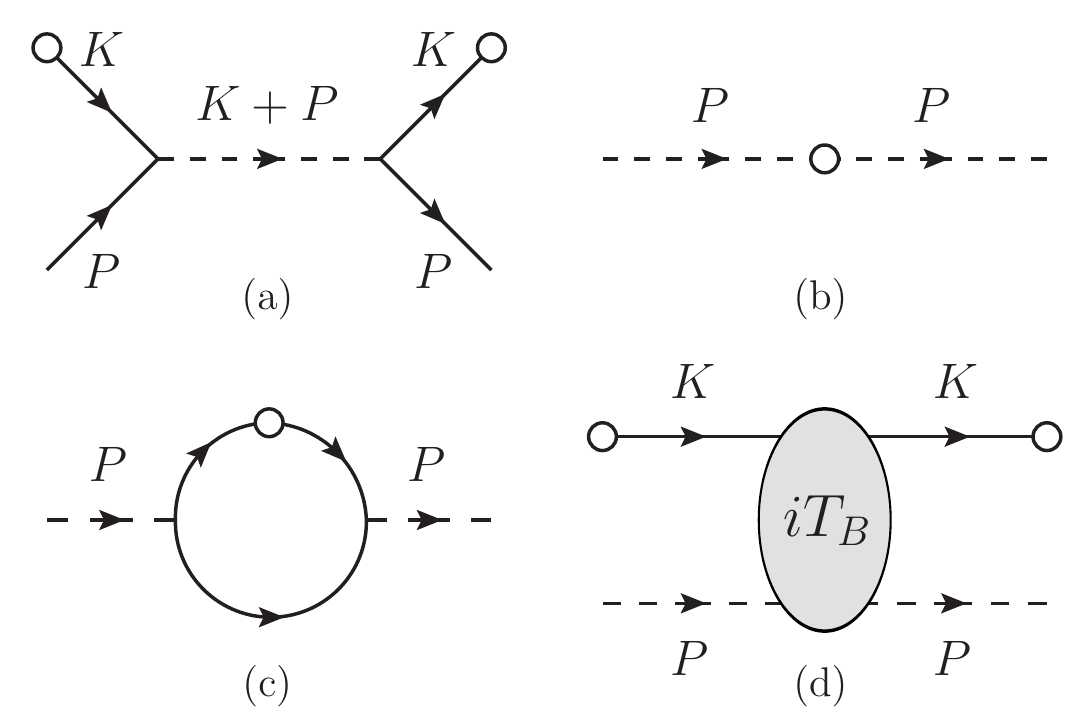}
\caption{\label{fig:Gphi}
Diagrams for the expectation values of (a) $\delta\G_\phi(K)=\G_\phi(K)-G(K)$ with respect to a one-boson state $|\phi_P\>$ and of (b) $\Phi^\+\Phi$, (c) $\O_{b,c}$, and (d) $\delta\G_\phi(K)$ with respect to a one-dimer state $|\Phi_P\>$.}
\end{figure}

Taking the vacuum expectation values of both sides of Eq.~(\ref{eq:Gphi_OPE}), we find
\begin{align}
W_\phi^1(K)=\<\mathrm{vac}|\G_\phi(K)|\mathrm{vac}\>=G(K).
\end{align}
Hereafter, $G(K)$ is subtracted from both sides of Eq.~(\ref{eq:Gphi_OPE}) and OPE for $\delta\G_\phi(K)=\G_\phi(K)-G(K)$ is considered, so that disconnected diagrams are canceled when its expectation values are evaluated.
The coefficients of $\O_{b,c}$ are derived by evaluating both sides of Eq.~(\ref{eq:Gphi_OPE}) with respect to a one-boson state $|\phi_P\>$.
The expectation values of $\O_{b,c}$ on the right-hand side are computed in Eq.~(\ref{eq:<1|One|1>_boson}).
On the other hand, the expectation value of $\delta\G_\phi(K)$ on the left-hand side is depicted in Fig.~\ref{fig:Gphi}(a) and is given by
\begin{align}\label{eq:<1|Gphi|1>_LHS}
\<\phi_P|\delta\G_\phi(K)|\phi_P\>=-A_B(K+P)[G(K)]^2.
\end{align}
We note that a diagram in which the two field operators are connected with different external lines without interactions is not considered because it contributes to $W_\phi^{\mathcal{O}_{b,c}}(K)$ only at $K=0$.
By comparing the expansion of $\<\phi_P|\delta\G_\phi(K)|\phi_P\>$ in $P$ with Eq.~(\ref{eq:<1|One|1>_boson}), the coefficients of $\O_{b,c}$ with $\Delta_{\O_{b,c}}\leq4$ are found to be
\begin{align}\label{eq:WphiphiOab(K)}
W_\phi^{\mathcal{O}_{b,c}}(K)=-\frac{1}{b!c!}\frac{\d^{b+c}A_B(K)}{\d\omega^b\d k^c}[G(K)]^2.
\end{align}
Unlike $W_{\hat{n}}^{\O_{b,c}}(K)$ [Eq.~(\ref{eq:WnnOab_in-general})] for the density correlation, $W_\phi^{\O_{b,c}}(K)$ is affected by the interaction through $A_B(K)$.
Therefore, a leading interaction effect on $\<\G_\phi(K)\>$ at large $K$ comes from the coefficient of $\O_{0,0}=\hat{n}$~\cite{Nishida:2012}.
This point is a characteristic of $\<\G_\phi(K)\>$ different from other correlation functions such as $S(K)$ and $\rho_B(k)$, whose asymptotic behaviors are governed by the two-body contact.

We now turn to the derivation of $W_\phi^{\Phi^\+\Phi}(K)$.
Unlike $S(K)$ and $\rho_B(k)$, whose coefficients of $\Phi^\+\Phi$ can be determined for any $K$ and $a$ within two-body calculations, we have to solve a three-body problem to compute $W_\phi^{\Phi^\+\Phi}(K)$.
Therefore, it is more difficult to determine than the coefficients for $S(K)$ and $\rho_B(k)$.
For convenience of the three-body calculation, we use a one-dimer state $|\Phi_P\>$ instead of a two-boson state used in Sec.~\ref{sec:density}.
First, we evaluate the expectation value of the right-hand side of Eq.~(\ref{eq:Gphi_OPE}) with respect to $|\Phi_P\>$.
The expectation values of $\Phi^\+\Phi$ and $\O_{b,c}$ can be expressed in terms of the Feynman diagrams as Figs.~\ref{fig:Gphi}(b) and \ref{fig:Gphi}(c), respectively.
The results are
\begin{align}
\<\Phi_P|\Phi^\+\Phi|\Phi_P\>&=1,\\
\<\Phi_P|\O_{b,c}|\Phi_P\>&=\frac{4}{m^2a^2}I_{b,c}(P),
\end{align}
where $I_{b,c}(P)$ is defined by Eq.~(\ref{eq:IOab}).
The expectation value of the right-hand side thus reads
\begin{align}\label{eq:<d|Gphi|d>_RHS}
&\sum_{\O\neq1}W_\phi^\O(K)\,\<\Phi_P|\O|\Phi_P\>\nonumber\\
&=\sum_{\Delta_{\O_{b,c}}\leq4}W_\phi^{\mathcal{O}_{b,c}}(K)\frac{4}{m^2a^2}I_{b,c}(P)
+W_\phi^{\Phi^\+\Phi}(K)+O(P).
\end{align}

On the other hand, the expectation value $\delta\G_\phi(K)$ on the left-hand side of Eq.~(\ref{eq:Gphi_OPE}) is given by the diagram in Fig.~\ref{fig:Gphi}(d), and it is evaluated as
\begin{align}\label{eq:<d|Gphi|d>_LHS}
\<\Phi_P|\delta\G_\phi(K)|\Phi_P\>=-[G(K)]^2T_B(K,P;K,P).
\end{align}
Here, $T_B(K,P;K',P')$ is the boson-dimer scattering amplitude where $K$ and $P$ ($K'$ and $P'$) are sets of initial (final) energy and momentum for a boson and a dimer, respectively.
Note $K+P=K'+P'$ because of the energy and momentum conservations.
Comparing the expansion of Eq.~(\ref{eq:<d|Gphi|d>_LHS}) in $P$ with Eq.~(\ref{eq:<d|Gphi|d>_RHS}), we obtain the following expression of $W_\phi^{\Phi^\+\Phi}(K)$:
\begin{align}\label{eq:W_phi^d^+d(K)}
W_\phi^{\Phi^\+\Phi}(K)&=\lim_{P\to0}\Bigg[-[G(K)]^2T_B(K,P;K,P)\nonumber\\
&\quad-\sum_{\Delta_{\O_{b,c}}\leq4}W_\phi^{\mathcal{O}_{b,c}}(K)\frac{4}{m^2a^2}I_{b,c}(P)\Bigg].
\end{align}
While Eqs.~(\ref{eq:I_O_bc}) show that $I_{b,c}(P)$ with small $\Delta_{\O_{b,c}}$ are divergent in $P\to0$, these divergences are exactly canceled by those from $T_B(K,P;K,P)$ in a similar way as in the 3D cases~\cite{Nishida:2012,Gubler:2015}.

\begin{figure}[t]
\includegraphics[width=\columnwidth,clip]{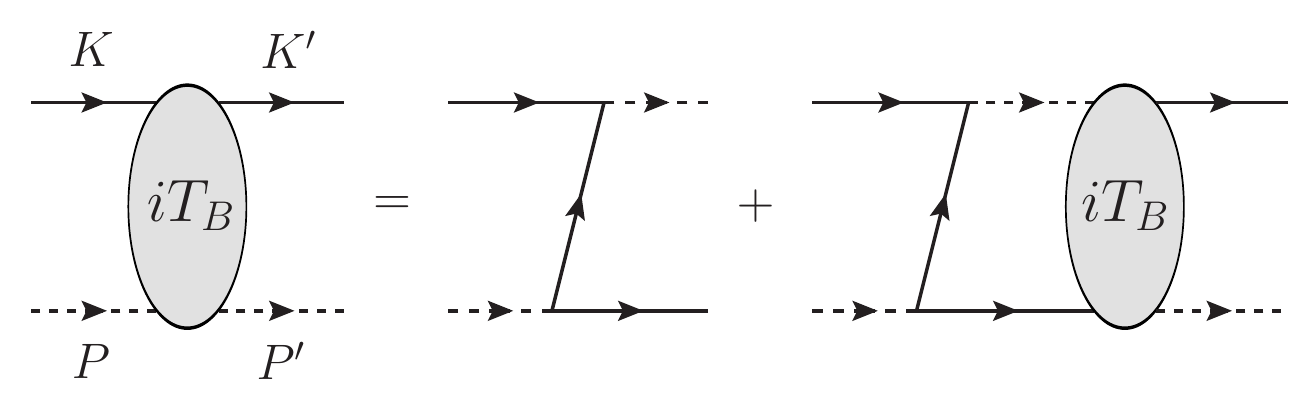}
\caption{\label{fig:T_B}Diagrammatic expression of the boson-dimer scattering amplitude [Eq.~(\ref{eq:T_B})].}
\end{figure}
The scattering amplitude solves the Skornyakov--Ter-Martirosyan (STM) equation depicted in Fig.~\ref{fig:T_B}:
\begin{align}\label{eq:T_B}
&T_B(K,P;K',P')\nonumber\\
&=t_B(K,P;K',P')-i\int_Qt_B(K,P;Q,K+P-Q)\nonumber\\
&\quad\times G(Q)D_B(K+P-Q)T_B(Q,K+P-Q;K',P'),
\end{align}
where the inhomogeneous term is given by
\begin{align}\label{eq:T^0_B}
t_B(K,P;K',P')=-\frac{4}{m^2a^2}G(P'-K).
\end{align}
One can solve this STM equation nonperturbatively by the numerical method used in the 3D cases~\cite{Nishida:2012,Gubler:2015}.
As explained previously, however, a perturbative calculation is available to determine the Wilson coefficients at large $K$ in the case of 1D bosons. 
For this reason, we evaluate $T_B(K,P;K',P')$ perturbatively in terms of $-1/a$ in this paper.
We then find $t_B(K,P;K',P')=O(a^{-2})$, while the loop corrections corresponding to the integral in Eq.~(\ref{eq:T_B}) make higher-order contributions.
In addition, Eq.~(\ref{eq:WphiphiOab(K)}) combined with Eq.~(\ref{eq:A_B(K)}) shows that the sum in Eq.~(\ref{eq:W_phi^d^+d(K)}) is $O(a^{-3})$.
Therefore, the large-$K$ asymptotics of $W_\phi^{\Phi^\+\Phi}(K)$ is found to be
\begin{align}\label{eq:W_phi^dd(K)}
W_\phi^{\Phi^\+\Phi}(K)=\frac{4}{m^2a^2}G(-K)[G(K)]^2+O(K^{-7}).
\end{align}
Note that power counting of the corrections in $a^{-1}$ combined with dimensional analysis leads to that in $K^{-1}$.

\subsection{\label{sec:A_B(K)}Single-particle spectral density}
Let us now consider the single-particle properties of 1D bosons in the large-$K$ limit.
First, we study quasiparticle energy and width near the single-particle peak $\omega\approx k^2/(2m)$.
The single-particle Green's function can be expanded in large $K$ by taking the thermal average of OPE in Eq.~(\ref{eq:Gphi_OPE}).
We consider the expansion up to $O(K^{-6})$.
By using Eqs.~(\ref{eq:WphiphiOab(K)}) and (\ref{eq:W_phi^dd(K)}) as well as $\<\O_{0,1}\>=0$ due to the inversion invariance of a given thermal state, $\<\G_\phi(K)\>$ reads
\begin{align}\label{eq:<Gphi(K)>_OPE}
\<\G_\phi(K)\>
=G(K)+W_\phi^{\hat{n}}(K)\,n+W_\phi^{\Phi^\+\Phi}(K)\,\mathcal{C}_2+O(K^{-7})
\end{align}
with $n=\<\hat{n}\>=\<\O_{0,0}\>$ and $\mathcal{C}_2=\<\Phi^\+\Phi\>$.
In general, the Green's function takes the form of $\<\G_\phi(K)\>=1/[G^{-1}(K)-\Sigma(K)]$ with the self-energy $\Sigma(K)$.
The self-energy up to $O(K^{-2})$ is thus given by
\begin{align}
\Sigma(K)
&=\frac{W_\phi^{\hat{n}}(K)}{[G(K)]^2}\,n+\frac{W_\phi^{\Phi^\+\Phi}(K)}{[G(K)]^2}\mathcal{C}_2+O(K^{-3})\nonumber\\
&=-\frac{4n}{ma}\left(1+\frac{1}{a\beta _K}+\frac{1}{(a\beta _K)^2}\right)+\frac{4\mathcal{C}_2\,G(-K)}{m^2a^2}\nonumber\\
&\quad+O(K^{-3}).
\end{align}

The pole $\omega=\omega_\mathrm{pole}$ of $\<\G_\phi(K)\>$ in a complex plane of $\omega$ gives the quasiparticle energy $\varepsilon_B(k)$ and width $\Gamma_B(k)$ as
\begin{align}
\varepsilon_B(k)=\Re[\omega_\mathrm{pole}],\qquad \Gamma_B(k)=-\Im[\omega_\mathrm{pole}].
\end{align}
Within our working accuracy, the pole near the single-particle peak is given by $\omega_\mathrm{pole}=k^2/(2m)+\Sigma(k^2/(2m),k)+O(k^{-3})$, while the quasiparticle residue is $\mathcal{Z}_B=1+O(k^{-3})$.
As a result, $\varepsilon_B(k)$ and $\Gamma_B(k)$ in the high-energy region are found to be
\begin{subequations}\label{eq:quasiparticle_bosons}
\begin{align}\label{eq:varepsilon_B(k)_result}
\varepsilon_B(k)
&=\frac{k^2}{2m}\Bigg[1+4\gamma\left(\frac{n}{k}\right)^2-2\gamma^2\left(2\gamma+\frac{\mathcal{C}_2}{n^2}\right)\left(\frac{n}{k}\right)^4\nonumber\\
&\qquad\qquad+O(k^{-5})\Bigg],\\
\label{eq:Gamma_B(k)_result}
\Gamma_B(k)
&=\frac{k^2}{2m}\Bigg[4\gamma^2\left(\frac{n}{|k|}\right)^3+O(k^{-5})\Bigg],
\end{align}
\end{subequations}
where $\gamma=-2/(na)$ is a dimensionless coupling constant used in the studies of 1D bosons.
The shift of $\varepsilon_B(k)$ from $k^2/(2m)$ depends on $n$ and $\mathcal{C}_2$, which is exactly calculable by the Bethe ansatz, while $\Gamma_B(k)$ depends only on $n$.
The effect of $\mathcal{C}_2$ on $\Gamma_B(k)$ is expected to come from loop corrections of $T_B(K,P;K,P)$ and to appear as subleading terms.
With Eqs.~(\ref{eq:quasiparticle_bosons}), the large-$K$ behavior of $\mathcal{A}_B(K)$ near $\omega\approx k^2/(2m)$ takes the form of
\begin{align}
\mathcal{A}_B(\omega,k)\simeq\frac1{\pi}\frac{\Gamma_B(k)}{[\omega-\varepsilon_B(k)]^2+[\Gamma_B(k)]^2}.
\end{align}
Equation~(\ref{eq:Gamma_B(k)_result}) shows that the width $\Gamma_B(k)\sim|k|^{-1}$ of the single-particle peak decreases with increasing $|k|$ as in the 3D cases with $s$-wave interactions~\cite{Nishida:2012}.
Note that the quasiparticle width in the 2D case decreases logarithmically with increasing the momentum.

We next turn to $\mathcal{A}_B(K)$ in the high-energy region away from the single-particle peak.
From Eq.~(\ref{eq:WphiphiOab(K)}), the imaginary part of the coefficient for $\hat{n}$ is given by
\begin{align}\label{eq:ImW^n}
&-\frac{1}{\pi}\Im\left[W_\phi^{\hat{n}}(K)\right]\nonumber\\
&=\theta(m\omega-k^2/4)\frac{4}{\pi m}[G(K)]^2
\frac{\sqrt{m\omega-k^2/4}}{1+a^2(m\omega-k^2/4)}.
\end{align}
This imaginary part is of the order of $O(K^{-5})$ for $\omega>k^2/(4m)$.
The Heaviside step function represents the two-particle threshold as in Eq.~(\ref{eq:Im[Wnnd^+d(K)]}).
On the other hand, Eq.~(\ref{eq:W_phi^dd(K)}) shows that the leading term of $W_\phi^{\Phi^\+\Phi}(K)$ is real away from the single-particle peak.
As a result, the large-$K$ behavior of $\mathcal{A}_B(K)$ for $\omega>k^2/(4m)$ is found to be proportional to $n$:
\begin{align}\label{eq:Aphi(K)_result}
\mathcal{A}_B(\omega,k)
=\frac{4mn}{\pi a^2}\frac{1}{\sqrt{m\omega-k^2/4}(m\omega-k^2/2)^2}.
\end{align}
This behavior holds when $\sqrt{m\omega}$ and $|k|$ are much larger than $n$, $1/|a|$, and $\sqrt{mT}$.

At the end of this subsection, we comment on $\mathcal{A}_B(K)$ for the Tonks-Girardeau gas with a hardcore repulsion.
In order to derive Eqs.~(\ref{eq:varepsilon_B(k)_result}), (\ref{eq:Gamma_B(k)_result}), and (\ref{eq:Aphi(K)_result}), we assumed that $\sqrt{m\omega}$ and $|k|$ are much larger than $|a|^{-1}$.
These results are thus not valid for the Tonks-Girardeau gas with $-1/a\to\infty$.
Nevertheless, OPE itself is available to study $\mathcal{A}_B(K)$ of the Tonks-Girardeau gas in the large-$K$ limit.
The one-body operators $\O_{b,c}$ with scaling dimensions $\Delta_{\O_{b,c}}=2b+c+1\leq4$ [see Eq.~(\ref{eq:Obc})] are well defined even in the case of $a\to-0$.
A well-defined auxiliary field for $a\to-0$ is given by $\tilde{\Phi}=a^{-1}\Phi$ and the corresponding dimer propagator is $i\tilde{D}_B(K)=im\beta_K$.
The dimer density operator $\tilde{\Phi}^\+\tilde{\Phi}$ has dimension $\Delta_{\tilde{\Phi}^\+\tilde{\Phi}}=4$.
The Wilson coefficients of $\O_{b,c}$ are obtained as Eq.~(\ref{eq:WphiphiOab(K)}) in the limit of $a\to-0$, leading to the large-$K$ behavior of $\mathcal{A}_B(K)$ for $\omega>k^2/(4m)$ given by
\begin{align}\label{eq:A_TG(K)}
\mathcal{A}_B(\omega,k)
=\frac{4mn}{\pi}\frac{\sqrt{m\omega-k^2/4}}{(m\omega-k^2/2)^2}.
\end{align}
On the other hand, the coefficient of $\tilde{\Phi}^\+\tilde{\Phi}$ equals $\lim_{a\to-0}[a^2W_\phi^{\Phi^\+\Phi}(K)]$ [see Eq.~(\ref{eq:W_phi^d^+d(K)})].
The determination of this coefficient requires a nonperturbative computation of $\tilde{T}_B(K,P;K',P')=\lim_{a\to-0}[a^2T_B(K,P;K',P')]$, which is beyond the scope of this paper.
Such a three-body calculation is expected to be performed by the method used in Refs.~\cite{Nishida:2012,Gubler:2015}.

\section{\label{sec:fermion}Fermions}
In this section, we study QFT for spinless fermions corresponding to bosons studied in Sec.~\ref{sec:boson}.
We consider the following Lagrangian density:
\begin{align}\label{eq:L_F}
\mathcal{L}_F&=\psi^\+\left(i\d_t+\frac{\d_x^2}{2m}\right)\psi-\frac{1}{mv_2}\Psi^\+\Psi\nonumber\\
&\quad+\frac{1}{m}\left[\Psi^\+\left(\psi(-i\tensor{\d}_x)\psi\right)+\left(\psi^\+(-i\tensor{\d}_x)\psi^\+\right)\Psi\right]\nonumber\\
&\quad+\frac{v_3}{m}\Psi^\+\psi^\+\psi\Psi.
\end{align}
Here, $\psi$ with dimension $\Delta_\psi=1/2$ is a fermionic field and $\Psi$ with dimension $\Delta_\Psi=1$ is an auxiliary bosonic field representing the degree of freedom of a dimer.
The coupling constant $v_2$ characterizes the coupling between two fermions.
When we focus on a two-fermion problem, we can neglect the last term in $\mathcal{L}_F$ and perform the path integrals over $\Psi$ and $\Psi^\+$, leading to the Lagrangian density with a local two-body interaction:
\begin{align}
\mathcal{L}_F'=\psi^\+\left(i\d_t+\frac{\d_x^2}{2m}\right)\psi+\frac{v_2}{m}\left|\psi(-i\tensor{\d}_x)\psi\right|^2.
\end{align}
This Lagrangian density is equivalent to the model considered in Ref.~\cite{Cui:2016a}.
By calculating a two-fermion scattering amplitude and matching it to the leading term in the effective-range expansion, the two-body sector can be regularized by renormalizing $v_2$.
On the other hand, the last term in Eq.~(\ref{eq:L_F}) provides the coupling between a fermion and a dimer and is not considered in the previous work.
This term involves a dimensionless coupling constant $v_3$ and represents a three-body coupling for fermions.
Since this term is marginal in the sense of the renormalization group, it should be taken into account in general.
As shown in the next subsection, $v_3\neq0$ plays a crucial role to regularize the three-body sector. 

We now present notations in Feynman diagrams.
The propagator of a fermion is equivalent to $G(K)$ in Eq.~(\ref{eq:G(K)}) and is also denoted by a solid line.
A dashed (dotted) line denotes a full (bare) propagator $iD_F(K)$ ($iD^{(0)}_F=-imv_2$) of a dimer.
A vertex where a dashed or dotted line is connected with two fermion lines is $2i/m$ multiplied by a relative momentum of fermions.
Solving the Dyson equation for $iD_F(K)$ given by the same diagram as for bosons [see Fig.~\ref{fig:A&D}(a)], we obtain
\begin{align}
D_F(K)=\frac{m}{1/a-\beta_K},
\end{align}
where $\beta_K=\sqrt{k^2/4-m\omega-i0^+}$ and the scattering length $a$ is related to $v_2$ and a momentum cutoff $\Lambda$ as
\begin{align}\label{eq:v2(Lambda)}
\frac{1}{v_2}=\frac{2\Lambda}{\pi}-\frac{1}{a}.
\end{align}
The two-fermion scattering amplitude $iA_F(K;P_1,P_1')$ is given by the diagram in Fig.~\ref{fig:A&D}(b) and equals
\begin{align}\label{eq:A_F}
A_F(K;P_1,P_1')=-\frac{4P_1P_1'}{m^2}D_F(K).
\end{align}
For fermions, the scattering amplitude depends not only on a total energy $\omega$ and a center-of-mass momentum $k$ but also on initial and final relative momenta $P_1$ and $P_1'$.

In this section, we proceed as follows:
The former half of Sec.~\ref{sec:three-body} is devoted to a scattering problem of a fermion and a dimer to determine $v_3$.
In order to confirm the validity of the obtained coupling constant, we calculate the binding energy of three fermions in the latter half and rederive the energy relation with a three-body contact in Sec.~\ref{sec:energy-relation}.
The asymptotic behaviors of dynamic correlation functions at large energy and momentum are discussed in Sec.~\ref{sec:correlations_femrions}.

\subsection{\label{sec:three-body}Three-body problem}
\begin{figure}[t]
\includegraphics[width=\columnwidth,clip]{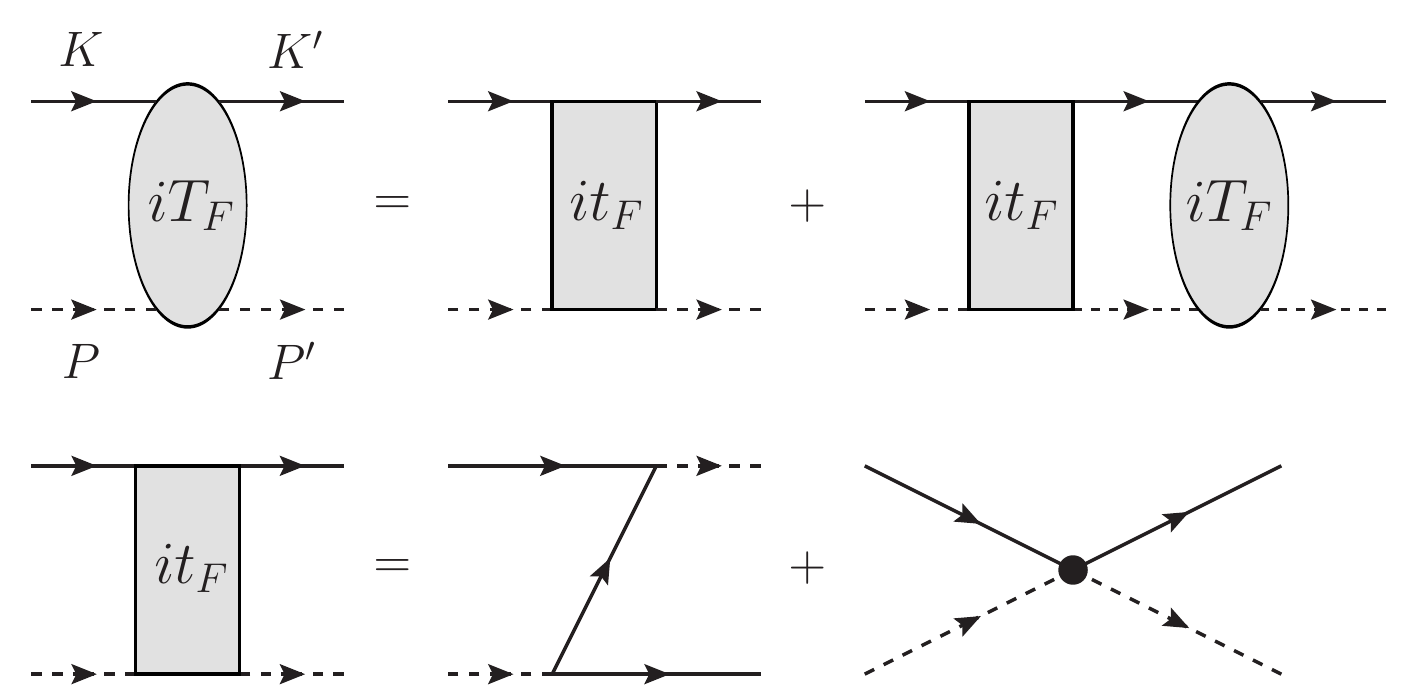}
\caption{\label{fig:T_F}
Feynman diagrams for the fermion-dimer scattering amplitude $iT_F$ and its tree terms $it_F$.
The solid and dashed lines indicate $iG(K)$ and $iD_F(K)$, respectively, while the dot denotes $iv_3/m$.}
\end{figure}

We start by considering the scattering problem of a fermion and a dimer, where the incoming fermion and dimer have sets of energy and momentum, $K$ and $P$, respectively, and the outgoing fermion and dimer have $K'$ and $P'$, respectively.
The fermion-dimer scattering amplitude $iT_F(K,P;K',P')$ solves the STM equation depicted in Fig.~\ref{fig:T_F}:
\begin{align}\label{eq:T_F}
&T_F(K,P;K',P')\nonumber\\
&=t_F(K,P;K',P')-i\int_Qt_F(K,P;Q,K+P-Q)\nonumber\\
&\quad\times G(Q)D_F(K+P-Q)T_F(Q,K+P-Q;K',P'),
\end{align}
where the inhomogeneous term is given by
\begin{align}
&t_F(K,P;K',P')\nonumber\\
&=(P_1-2K'_1)(P'_1-2K_1)\frac{G(P-K')}{m^2}+\frac{v_3}m.
\end{align}
Note $K+P=K'+P'$ because of the energy and momentum conservations.
The integrand in Eq.~(\ref{eq:T_F}) has only one pole $Q_0=Q_1^2/(2m)-i0^+$ in the lower half-plane of $Q_0$.
By performing the integration over $Q_0$, Eq.~(\ref{eq:T_F}) reads
\begin{align}
&T_F(K,P;K',P')\nonumber\\
&=t_F(K,P;K',P')-\int\frac{dQ_1}{2\pi}t_F(K,P;Q,K+P-Q)\nonumber\\
&\quad\times D_F(K+P-Q)T_F(Q,K+P-Q;K',P')|_{Q_0=\frac{Q_1^2}{2m}}.
\end{align}

This integral equation reduces to a simpler form under the on-shell condition in the center-of-mass frame.
Taking $K=\left(k^2/(2m),k\right)$, $K'=\left(k'^2/(2m),k'\right)$, and $K+P=K'+P'=(E,0)$, we obtain the equation for the on-shell scattering amplitude $T_F(k;k')$:
\begin{align}\label{eq:T_F(k:k')}
T_F(k;k')=t_F(k;k')-\int\frac{dq}{2\pi}t_F(k;q)D_F(q)T_F(q;k'),
\end{align}
where $Q_1\to q$, $D_F(q)=D_F(E-q^2/(2m),q)$, and
\begin{align}
mt_F(k;k')=\frac{2mE+3kk'}{mE+i0^+-(k^2+kk'+k'^2)}+v_3-2.
\end{align}
The integral in Eq.~(\ref{eq:T_F(k:k')}) for $|q|<\Lambda$ has ultraviolet divergences $\sim\ln\Lambda$ unless $v_3\to2$, where $t_F(k;k')$ decays by power law for large $k'$ with $E$ and $k$ fixed.
Usually, such a divergence is canceled by making a coupling constant dependent on $\Lambda$.
When the coupling constant is dimensionless, a new length scale emerges as a consequence of the dimensional transmutation~\cite{Coleman:1973}.
In nonrelativistic QFT, the dimensional transmutation is discussed in the 1D~\cite{Sekino:2018b,Drut:2018,Daza:2019,Camblog:2019} and 2D~\cite{Bergman:1992} cases.
However, since the 1D scattering length $a$ is only the length scale associated with the contact interaction, the emergence of an additional scale is prohibited in our case.
Therefore, we conclude that the three-body coupling constant must be
\begin{align}\label{eq:v3}
v_3=2
\end{align}
so that the logarithmic divergences disappear without generating an additional scale.\footnote{As shown in Appendix~\ref{appendix:a_3}, fermions with a three-body attraction can be described by choosing $v_3=2+\pi/[2\sqrt3\ln(\sqrt{3}\Lambda a_3)]$.
Here, the emergent length scale $a_3$ generates the binding energy $E=-1/(ma_3^2)$ of three fermions in the limit of $a\to\infty$, which corresponds to a three-boson bound state without two-body but with three-body interactions~\cite{Sekino:2018b}.}

To confirm that the theory with Eq.~(\ref{eq:v3}) corresponds to the bosonic one studied in the previous section, we investigate a three-fermion bound state for $a>0$ and compute its binding energy.
If there is a three-body bound state with $E=-\kappa^2/m$, the fermion-dimer scattering amplitude in the limit of $E\to-\kappa^2/m$ takes the form of $T_F(k;k')\to Z_F(k)Z_F^*(k')/(E+\kappa^2/m)$.
Comparing residues of both sides of Eq.~(\ref{eq:T_F(k:k')}) with respect to $E=-\kappa^2/m$, we obtain the homogeneous integral equation for $z_F(k)\equiv Z_F(k)D_F(k)$:
\begin{align}\label{eq:z_F(k)}
&\left(\sqrt{\frac34k^2+\kappa^2}-\frac{1}{a}\right)z_F(k)\nonumber\\
&=\int\frac{dq}{2\pi}\frac{2\kappa^2-3kq}{k^2+kq+q^2+\kappa^2}z_F(q).
\end{align}
We can analytically obtain one solution $z_F(k)=1/[\left(ka/2\right)^2+1]$ with $\kappa=2/a$.
This bound state has the binding energy $E=-4/(ma^2)$, which is identical to that of a three-boson bound state found by McGuire~\cite{McGuire:1964} and thus confirms Eq.~(\ref{eq:v3}).
We next rederive the energy relation as another demonstration of the validity of Eq.~(\ref{eq:v3}).

\subsection{\label{sec:energy-relation}Contacts and the energy relation}
Before turning to the energy relation, we derive the expressions of the two- and three-body contact densities in terms of field operators.
By recalling Eqs.~(\ref{eq:contacts}), the contact densities are given by
\begin{subequations}\label{eq:contacts_fermion}
\begin{align}
\mathcal{C}_2&=\lim_{y\to x}\<\hat{n}(t,x)\hat{n}(t,y)\>,\\
\label{eq:C3}
\mathcal{C}_3&=\lim_{y,z\to x}\<\hat{n}(t,x)\hat{n}(t,y)\hat{n}(t,z)\>
\end{align}
\end{subequations}
in the Heisenberg picture.
In the fermionic theory, the number density operator is given by $\hat{n}=\psi^\+\psi$.
One may think that $\mathcal{C}_2$ and $\mathcal{C}_3$ vanish due to $\psi^2=0$ resulting from the Fermi statistics.
However, the presence of the contact interaction leads to the renormalization of composite operators, which is encoded in $\Psi$, and thus $\mathcal{C}_2$ and $\mathcal{C}_3$ take nonzero values~\cite{Cui:2016a,Sekino:2018a}.

To obtain the explicit forms of $\mathcal{C}_2$ and $\mathcal{C}_3$, we evaluate an equal-time OPE:
\begin{align}\label{eq:OPE_nn}
\hat{n}(t,x)\hat{n}(t,y)=\sum_\O w^\O(x-y)\O(t,x).
\end{align}
As mentioned previously, $\psi$ and $\Psi$ have dimensions $\Delta_\psi=1/2$ and $\Delta_\Psi=1$, respectively.
Under the equal-time condition, the unit and one-body operators have vanishing coefficients because matrix elements of $\hat{n}(t,x)\hat{n}(t,y)$ are zero in the vacuum and in the one-fermion sector.
As a result, the lowest-order local operator whose coefficient takes a nonzero value is $\O=\Psi^\+\Psi$ with $\Delta_{\Psi^\+\Psi}=2$.
By dimensional analysis, we see that coefficients of local operators with larger scaling dimensions vanish in the limit of $x-y\to0$.

\begin{figure}[t]
\includegraphics[width=0.6\columnwidth,clip]{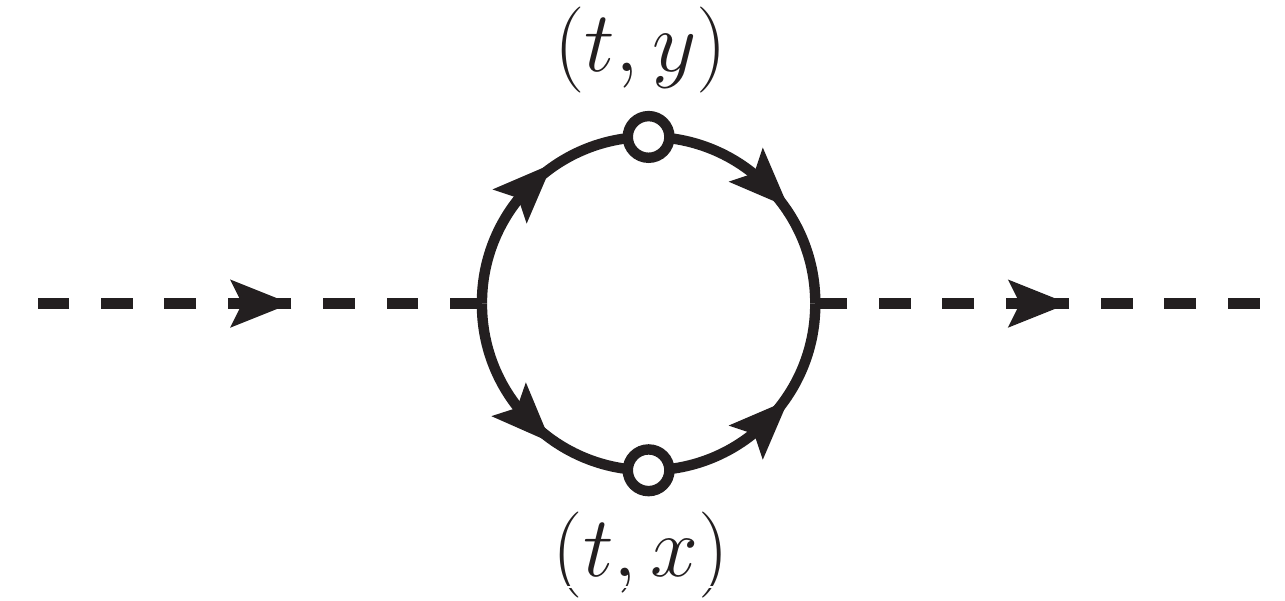}
\caption{\label{fig:n(X)n(Y)}
Diagram for the expectation value of $\hat{n}(t,x)\hat{n}(t,y)$ with respect to a one-dimer state.}
\end{figure}
In order to determine $w^{\Psi^\+\Psi}(x-y)$, we employ the matching procedure with respect to a one-dimer state $|\Psi_P\>$.
The expectation value of $\Psi^\+\Psi(t,x)$ on the right-hand side of Eq.~(\ref{eq:OPE_nn}) is given by the diagram in Fig.~\ref{fig:Gphi}(b) and equals
\begin{align}\label{eq:<|varphi^+varphi|>}
\<\Psi_P|\Psi^\+\Psi(t,x)|\Psi_P\>=1.
\end{align}
On the other hand, the expectation value of the left-hand side of Eq.~(\ref{eq:OPE_nn}) is depicted in Fig.~\ref{fig:n(X)n(Y)} and equals
\begin{align}
&\<\Psi_P|\hat{n}(t,x)\hat{n}(t,y)|\Psi_P\>\nonumber\\
&=4\int\frac{dq}{2\pi}\frac{q\,e^{iq(x-y)}}{q^2+\beta_P^2}\int\frac{dq'}{2\pi}\frac{q	'e^{-iq'(x-y)}}{q'^2+\beta_P^2}.
\end{align}
Performing the integrations by the residue theorem yields
$\<\Psi_P|\hat{n}(t,x)\hat{n}(t,y)|\Psi_P\>=e^{-2\beta_P|x-y|}$.
By comparing this expectation value for $x-y\to0$ with Eq.~(\ref{eq:<|varphi^+varphi|>}), the coefficient of $\Psi^\+\Psi$ is found to be unity, leading to
\begin{align}\label{eq:OPE_nn2}
\hat{n}(t,x)\hat{n}(t,y)=\Psi^\+\Psi(t,x)+O(x-y).
\end{align}
Substituting this operator relation into Eqs.~(\ref{eq:contacts_fermion}), we obtain
\begin{align}\label{eq:C_2F}
\mathcal{C}_2&=\<\Psi^\+\Psi(t,x)\>,\\
\label{eq:C_3F_0}
\mathcal{C}_3&=\lim_{z\to x}\<\Psi^\+\Psi(t,x)\hat{n}(t,z)\>.
\end{align}
The limit in the second line can be taken by evaluating OPE for $\Psi^\+\Psi(t,x)\hat{n}(t,z)$ in a similar way as for Eq.~(\ref{eq:OPE_nn2}).
As a result, the three-body contact density is found to be
\begin{align}\label{eq:C_3F}
\mathcal{C}_3=\<\Psi^\+\psi^\+\psi\Psi(t,x)\>.
\end{align}

We now rederive the energy relation for fermions [Eq.~(\ref{eq:energy_fermions})].
The energy for a thermal state is obtained as $E=\int\!dx\,\<\mathcal{H}_F\>$, where the Hamiltonian density of the system is given by
\begin{align}
\mathcal{H}_F
&=\frac{|\d_x \psi|^2}{2m}+\frac{1}{mv_2}\Psi^\+\Psi-\frac{v_3}{m}\Psi^\+\psi^\+\psi\Psi\nonumber\\
&-\frac{1}{m}\left[\Psi^\+\left(\psi(-i\tensor{\d}_x)\psi\right)+\left(\psi^\+(-i\tensor{\d}_x)\psi^\+\right)\Psi\right].
\end{align}
By using the Euler-Lagrange equations for $\Psi$ and $\Psi^\+$,
\begin{align}
\psi(-i\tensor{\d}_x)\psi=\frac{1}{v_2}\Psi-v_3\psi^\+\psi\Psi,
\end{align}
the thermal average of the second line reduces to
\begin{align}
&-\frac{1}{m}\left\<\Psi^\+\left(\psi(-i\tensor{\d}_x)\psi\right)+\left(\psi^\+(-i\tensor{\d}_x)\psi^\+\right)\Psi\right\>\nonumber\\
&=-\frac{2}{mv_2}\<\Psi^\+\Psi\>+\frac{2v_3}{m}\<\Psi^\+\psi^\+\psi\Psi\>
\end{align}
so that the energy reads
\begin{align}
E=\int\!dx\left[\frac{\<|\d_x \psi|^2\>}{2m}
-\frac{\<\Psi^\+\Psi\>}{mv_2}+\frac{v_3}{m}\<\Psi^\+\psi^\+\psi\Psi\>\right].
\end{align}
From Eqs.~(\ref{eq:C_2F}) and (\ref{eq:C_3F}), the second and third terms in the integrand are proportional to $\mathcal{C}_2$ and $\mathcal{C}_3$, respectively.
Substituting the explicit forms of $v_2$ and $v_3$ [Eqs.~(\ref{eq:v2(Lambda)}) and (\ref{eq:v3})] into this, the energy relation is found to be
\begin{align}\label{eq:E_F_QFT}
E=\int\frac{dk}{2\pi}\frac{k^2}{2m}\left(\rho_F(k)-\frac{4C_2}{k^2}\right)+\frac{C_2}{ma}+\frac{2C_3}{m},
\end{align}
where $\rho_F(k)=L\int\!dx\,e^{-ikx}\<\psi^\+(t,x)\psi(t,0)\>$ with $L$ being the system size is the momentum distribution in terms of field operators.
In the above derivation, the nonzero three-body coupling constant leads to the emergence of $C_3$ in the energy relation in agreement with Ref.~\cite{Sekino:2018a}.

\subsection{\label{sec:correlations_femrions}Single-particle spectral density}
This subsection is devoted to deriving the behaviors of dynamic correlation functions for fermions at large energy and momentum.
As mentioned previously, the dynamic structure factor $S(K)$ for fermions is identical to that for bosons.
Indeed, the asymptotic behaviors of $S(K)$ in Eqs.~(\ref{eq:S(K)_result}) and (\ref{eq:S(K)_near_TG}) can be rederived in the same way as for bosons.

Unlike $S(K)$, the single-particle spectral density $\mathcal{A}_F(K)$ for fermions shows large-$K$ behaviors different from those of $\mathcal{A}_B(K)$.
In terms of a time-ordered Green's function, $\mathcal{A}_F(K)$ is given by
\begin{align}\label{eq:Apsi(K)}
\mathcal{A}_F(K)=-\frac{1}{\pi}\Im[\<\G_\psi(K)\>]+O(e^{-\omega/T}).
\end{align}
In order to study $\mathcal{A}_F(K)$ at large $K$, we employ OPE:
\begin{align}\label{eq:Gpsi_OPE}
\G_\psi(K)=\sum_{\O}W_\psi^\O(K)\O.
\end{align}
In the fermionic theory, local operators with small scaling dimensions are the unit operator $1$ with $\Delta_1=0$, $\hat{n}$ with $\Delta_{\hat{n}}=1$, and $\hat{\mathcal{C}}_2=\Psi^\+\Psi$ with $\Delta_{\hat{\mathcal{C}}_2}=2$.
By the matching procedure in the vacuum and in the one-fermion sector, we can obtain Wilson coefficients of $1$ and $\hat{n}$ in the same way as for bosons.
On the other hand, a perturbative calculation of the three-body scattering amplitude used for bosons cannot be applied to the determination of $W_\psi^{\hat{\mathcal{C}}_2}(K)$ at large $K$.
As explained in Sec.~\ref{sec:OPE}, this is because the limit of $a\to\infty$ does not correspond to a weakly interacting limit in the case of fermions.
The STM equation in Eq.~(\ref{eq:T_F}) is expected to be nonperturbatively solved by the method used in Refs.~\cite{Nishida:2012,Gubler:2015}.
Since we focus on analytical studies in this paper, we consider OPE in Eq.~(\ref{eq:Gpsi_OPE}) up to dimension $\Delta_\O=1$.

As a result of the matching procedure, the single-particle Green's function reads
\begin{align}\label{eq:<G_psi(K)>}
\<\G_\psi(K)\>=G(K)+\frac{n}{m}\frac{k^2[G(K)]^2}{1/a-\beta_K}+O(K^{-4}).
\end{align}
We first consider the high-energy region near the single-particle peak $\omega\approx k^2/(2m)$.
Within our working accuracy, the quasiparticle energy and residue are not affected by the interaction; $\varepsilon_F(k)=k^2/(2m)+O(1)$ and $\mathcal{Z}_F=1+O(k^{-1})$.
On the other hand, the quasiparticle width is given by
\begin{align}
\label{eq:Gamma_F(k)_result}
\Gamma_F(k)=\frac{4n|k|}{m}+O(1).
\end{align}
While $\Gamma_F(k)$ grows with increasing $|k|$, the quasiparticle picture is still valid because of $\Gamma_F(k)/\varepsilon_F(k)\propto|k|^{-1}$.
This linear behavior of $\Gamma_F(k)$ results from both one-dimensionality and the strong interaction at $a\to\infty$.
Indeed, the quasiparticle width decreases with increasing the momentum for 1D bosons [see Eq.~(\ref{eq:Gamma_B(k)_result})] as well as in higher dimensions~\cite{Nishida:2012}.
The Wilson coefficient of $\hat{\mathcal{C}}_2$ provides $O(1)$ corrections in $\varepsilon_F(k)$ and $\Gamma_F(k)$ and a leading correction in $\mathcal{Z}_F$.

We next turn to $\mathcal{A}_F(K)$ in the high-energy region away from the single-particle peak.
Within our working accuracy, the imaginary part of $\<\G_\psi(K)\>$ in Eq.~(\ref{eq:<G_psi(K)>}) arises from $\beta_K$, which is pure imaginary only above the two-particle threshold.
As a result, the behavior of $\mathcal{A}_F(K)$ for $\omega>k^2/(4m)$ is obtained as
\begin{align}\label{eq:Apsi(K)_result}
\mathcal{A}_F(\omega,k)=\frac{mn}{\pi}\frac{k^2}{(m\omega-k^2/4)^{3/2}(m\omega-k^2/2)^2}.
\end{align}
This behavior holds when $\sqrt{m\omega}$ and $|k|$ are much larger than $n$, $1/|a|$, and $\sqrt{mT}$.
We note that the power-law tail does not appear in the limit of $a\to-0$, i.e., a noninteracting limit for fermions.
Indeed, the second term in Eq.~(\ref{eq:<G_psi(K)>}) vanishes in this limit.

\section{\label{sec:conclusion}Conclusion}
In this paper, we elucidated universal relations for 1D bosons and fermions related to each other via the Bose-Fermi mapping [Eq.~(\ref{eq:mapping})] from the viewpoint of QFT.
These universal relations are crucial properties of the systems because they are exact even in the strongly interacting regimes.
By taking advantage of OPE in the QFT formalism, high-energy behaviors of dynamic correlation functions [Eqs.~(\ref{eq:S(K)_result}), (\ref{eq:S(K)_near_TG}), (\ref{eq:quasiparticle_bosons}), (\ref{eq:Aphi(K)_result}), (\ref{eq:A_TG(K)}), (\ref{eq:Gamma_F(k)_result}), and (\ref{eq:Apsi(K)_result})] were derived.
While the dynamic structure factor is identical between bosons and fermions, the single-particle spectral densities differ between them.
In particular, we found that the sharpening (broadening) of the single-particle peak for bosons (fermions) results from the fact that $a\to\infty$ corresponds to a weakly (strongly) interacting limit.
The energy relation for fermions [Eq.~(\ref{eq:E_F_QFT})] was also rederived, where the emergence of the three-body contact was found to originate from a three-body coupling term with Eq.~(\ref{eq:v3}) in the QFT formalism.

Our results presented in this paper can be generalized in various directions.
The universal relations can be extended to the presence of effective-range corrections~\cite{Gurarie:2006,Imambekov:2010,Qi:2013}.
Indeed, the impact of such corrections on some universal relations has been studied~in 1D~\cite{Cui:2016b} as well as in higher dimensions~\cite{Braaten:2008b,Werner:2012}.
Another interesting extension is QFT for fermions in the presence of multi-body resonances.
In the case of 1D bosons, such resonances have been described by introducing higher-body interactions~\cite{Nishida:2010,Sekino:2018b,Nishida:2018,Pricoupenko:2018,Guijarro:2018}.
A three-boson attraction with dimensional transmutation leads to the formations of quantum droplet states~\cite{Sekino:2018b} and of excited few-body bound states~\cite{Nishida:2018,Pricoupenko:2018,Guijarro:2018}.
A resonant four-boson interaction results in the formation of Efimov pentamers in 1D~\cite{Nishida:2010}.
According to the Bose-Fermi mapping, these phenomena should also emerge for fermions.
As shown in Appendix~\ref{appendix:a_3}, the counterpart of the three-boson attraction can be introduced to fermions by considering the logarithmic running of the fermion-dimer coupling $v_3$ in our Lagrangian density [Eq.~(\ref{eq:L_F})].
The resonant four-fermion interaction leading to the Efimov effect for five fermions is expected to be described by adding $v_4\Psi^\+\Psi^\+\Psi\Psi$ to Eq.~(\ref{eq:L_F}) and tuning the dimer-dimer coupling $v_4$.

We note that, when this paper was being finalized, there appeared preprints~\cite{Valiente:2020a,Valiente:2020b} where the Bose-Fermi correspondence was generalized to arbitrary spin, single-particle dispersion, and low-energy interactions in the universal regime by using the effective field theory.
In particular, fermions corresponding to bosons with a three-body repulsion~\cite{Pastukhov:2019,Valiente:2019a,Valiente:2019b} were considered.
The binding energy of the three-fermion bound state, which we analytically obtained from Eq.~(\ref{eq:z_F(k)}), was also investigated.

\acknowledgments
The authors thank D.~Petrov and M.~Valiente for valuable discussions on their related works.
This work was supported by JSPS KAKENHI Grants No.\ JP19J01006 and No.\ JP18H05405.

\appendix
\section{\label{appendix:integrals}Loop integrals}
Here, the integrals corresponding to loops in Figs.~\ref{fig:<2|O|2>} and \ref{fig:<2|G_n|2>} are computed.
First, we calculate the integrals in Eq.~(\ref{eq:IOab}):
\begin{align}
I_{b,c}(P)=i\int_QG(Q)[G(P-Q)]^2(P_0-Q_0)^b(P_1-Q_1)^c,
\end{align}
where nonnegative integers $b,c$ are restricted to $\Delta_{\O_{b,c}}=2b+c+1\leq4$.
The integration can be performed by the residue theorem and the explicit forms of $I_{b,c}(P)$ are found to be
\begin{subequations}\label{eq:I_O_bc}
\begin{align}
I_{0,0}(P)&=\frac{m^2}{4\beta_P^3},\\
I_{0,1}(P)&=\frac{m^2}{8\beta_P^3}P_1,\\
I_{0,2}(P)&=\frac{m^2}{4\beta_P^3}(P_1^2/4+\beta_P^2),\\
I_{0,3}(P)&=\frac{m^2}{32\beta_P^3}P_1(P_1^2+12\beta_P^2),\\
I_{1,0}(P)&=\frac{m }{32 \beta_P^3}(P_1^2-12\beta_P^2),\\
I_{1,1}(P)&=\frac{m}{16\beta_P^3}P_1\left(P_1^2/4-\beta_P^2\right).
\end{align}
\end{subequations}

We next turn to the integrals in Eqs.~(\ref{eq:Js_def}):
\begin{subequations}
\begin{align}
J_1(K,P)&=i\int_QG(Q)G(P-Q)G(K+P-Q),\\
J_2(K,P)&=-i\int_QG(Q)G(Q+K)G(P-Q)\nonumber\\
&\quad\times G(P-K-Q),\\
J_3(K,P)&=i\int_QG(Q)[G(P-Q)]^2G(P+K-Q).
\end{align}
\end{subequations}
The analytical expressions of these integrals are found to be
\begin{widetext}
\begin{subequations}\label{eq:Js}
\begin{align}
J_1(K,P)&=\left(\frac{m^2}{2\beta _{K+P}}+\frac{m^2}{2\beta _P}\right)\frac{1}{\left(\beta _{K+P}+\beta _P\right)^2+k^2/4},\\
J_2(K,P)&=\frac{m^3}{2\beta_P}\left(\frac{1}{(k+i\beta_P)^2+\beta_P^2}\frac{1}{(k/2+i\beta_P)^2+\beta_{K+P}^2}+(\beta_P\to-\beta_P)\right)\nonumber\\
&\quad+\frac{m^3}{2\beta_{K+P}}\frac{1}{(k/2+i\beta_{K+P})^2+\beta_P^2}\frac{1}{(k/2-i\beta_{K+P})^2+\beta_P^2}
+(K\to-K),\\
J_3(K,P)&=\frac{m^3}{4\beta_P^3}\frac{2ik\beta _P-\beta _{K+P}^2-k^2/4+3\beta _P^2}{(ik\beta _P-\beta _{K+P}^2-k^2/4+\beta _P^2)^2}-\frac{m^3}{2\beta_{K+P}}\frac{1}{(-ik\beta _{K+P}+\beta _{K+P}^2-k^2/4-\beta _P^2)^2}.
\end{align}
\end{subequations}
Their expansions in $P$ yield
\begin{subequations}\label{eq:Js_exp}
\begin{align}
J_1(K,P)&=-\frac{m}{2\beta _P}\left(G(K)+\frac{kP_1[G(K)]^2}{2m}\right)-\beta_K[G(K)]^2-\frac{mG(K)}{2\beta _{K}}+O(P),\\
J_2(K,P)&=\frac{m}{\beta_P}\left.\left.\left(G(K)+\frac{kP_1[G(K)]^2}{2m}\right)\right(K\to-K\right)
+\frac{m[G(K)]^2}{2\beta_{K}}+\frac{m[G(-K)]^2}{2\beta_{-K}}+O(P),
\\
J_3(K,P)&=\frac{[G(K)]^4}{2m}\left(k^2\beta_K-\frac{(m\omega)^2}{\beta_K}\right)
+\sum_{\Delta_{\O_{b,c}}\leq4}\frac{1}{b!c!}\frac{\d^{b+c}G(K)}{\d\omega^b\d k^c}I_{b,c}(P)+O(P).
\end{align}
\end{subequations}
\end{widetext}

\section{\label{appendix:a_3}Fermions with a three-body attraction}
In Sec.~\ref{sec:fermion}, the three-body coupling constant was fixed as $v_3=2$ [Eq.~(\ref{eq:v3})] to regularize the three-body sector for fermions without generating an additional scale.
Here, we consider another possibility for the regularization, i.e., $v_3\to2$ depending logarithmically on a momentum cutoff $\Lambda$.
Such a $v_3$ describes a three-body attraction characterized by the three-body scattering length $a_3$.
In what follows, we study bound states of three fermions with the three-body attraction.

An integral equation for three-fermion bound states can be derived from the STM equation [Eq.~(\ref{eq:T_F(k:k')})] in a similar way as for Eq.~(\ref{eq:z_F(k)}).
If there is a three-body bound state with $E=-\kappa^2/m$, the fermion-dimer scattering amplitude in the limit of $E\to-\kappa^2/m$ takes the form of $T_F(k;k')\to Z_F(k)Z_F^*(k')/(E+\kappa^2/m)$.
Comparing residues of both sides of Eq.~(\ref{eq:T_F(k:k')}) with respect to $E=-\kappa^2/m$, we obtain the homogeneous integral equation for $z_F(k)\equiv Z_F(k)D_F(k)$:
\begin{align}\label{eq:z_F(k)_appendix}
&\left(a-\frac{1}{\sqrt{\frac34k^2+\kappa^2}}\right)z_F(k)\nonumber\\
&=\frac{a}{\sqrt{\frac34k^2+\kappa^2}}\int\frac{dq}{2\pi}\frac{2\kappa^2-3kq}{k^2+kq+q^2+\kappa^2}z_F(q)\nonumber\\
&\quad+\frac{w_F}{\sqrt{\frac34k^2+\kappa^2}}
\end{align}
with $w_F=a(v_3-2)\int dq\,z_F(q)/(2\pi)$.
Because the integration of both sides over $k$ leads to
\begin{align}
&\frac{w_F}{v_3-2}-\int\frac{dk}{2\pi}\frac{z_F(k)}{\sqrt{\frac34k^2+\kappa^2}}\nonumber\\
&=\frac{2}{3}\frac{w_F}{v_3-2}+\frac{2w_F}{\sqrt3\pi}\ln(\sqrt3\Lambda/\kappa),
\end{align}
we can find
\begin{align}\label{eq:w_F}
w_F=\frac{\sqrt{3}\pi}{2\ln(a_3\kappa)}\int\frac{dq}{2\pi}\frac{z_F(q)}{\sqrt{\frac34q^2+\kappa^2}},
\end{align}
where an emergent length scale $a_3>0$ is introduced by
\begin{align}\label{eq:v3_appendix}
\frac{1}{v_3-2}=\frac{2\sqrt3}{\pi}\ln(\sqrt3\Lambda a_3).
\end{align}
The three-fermion bound states are obtained by solving Eq.~(\ref{eq:z_F(k)_appendix}) with (\ref{eq:w_F}).

The above three-fermion bound states correspond to three-boson bound states with two- and three-body interactions~\cite{Nishida:2018,Pricoupenko:2018,Guijarro:2018}.
The integral equation for such bosons is given by
\begin{align}\label{eq:z_B(k)}
&\left(a-\frac{1}{\sqrt{\frac34k^2+\kappa^2}}\right)z_B(k)\nonumber\\
&=\int\frac{dq}{2\pi}\frac{4z_B(q)}{k^2+kq+q^2+\kappa^2}+\frac{3w_B}{\sqrt{\frac34k^2+\kappa^2}}
\end{align}
with
\begin{align}\label{eq:w_B}
w_B=\frac{\sqrt{3}\pi}{2\ln(a_3\kappa)}\int\frac{dq}{2\pi}\frac{z_B(q)}{\sqrt{\frac34q^2+\kappa^2}},
\end{align}
which was analytically solved in Ref.~\cite{Guijarro:2018}.\footnote{Our definition of $a_3$ is consistent with that in Ref.~\cite{Nishida:2018} but different from that in Ref.~\cite{Guijarro:2018} by a factor $e^\gamma/2$ with Euler's constant $\gamma\approx0.577$.}
Employing the same method as for bosons, we can analytically solve Eq.~(\ref{eq:z_F(k)_appendix}) for fermions and find that the solutions are identical between bosons and fermions.
The solutions of Eqs.~(\ref{eq:z_B(k)}) and (\ref{eq:z_F(k)_appendix}) are both given by
\begin{align}
z_{B/F}(k)&=\frac{2w_{B/F}}{\pi}\int_0^\infty\!\!dp\,\frac{\sqrt{3p^2+4\kappa^2}}{2-a\sqrt{3p^2+4\kappa^2}}\frac{f_p(k)}{p^2+\kappa^2}\nonumber\\
&\quad+\frac{4w_{B/F}}{a\kappa-2}\frac{\kappa^2}{k^2+\kappa^2},
\end{align}
where
\begin{align}
f_p(k)&=
\frac{\pi\kappa^2\delta(k+p)}{\sqrt{3p^2+4\kappa^2}}+\mathcal{P}\frac{p}{k+p}-\frac{p^2+\kappa^2+\frac{k^2p^2}{k^2+p^2+\kappa^2}}{k^2+kp+p^2+\kappa^2}\nonumber\\
&\quad+(k\to-k)
\end{align}
and $\mathcal{P}$ denotes the Cauchy principal value.
Substituting the above solutions into Eqs.~(\ref{eq:w_B}) and (\ref{eq:w_F}) yields the same equation to determine $\kappa$:
\begin{align}
\ln(a_3\kappa)=\frac{1}{(a\kappa)^2-4}\left\{\frac{8\pi}{3\sqrt3}+[3(a\kappa)^2-4]g(a\kappa)\right\}.
\end{align}
Here,
\begin{align}
g(a\kappa)=-\frac{\ln\left[\frac{1+\sqrt{1-(a\kappa)^2}}{1-\sqrt{1-(a\kappa)^2}}\right]}{2\sqrt{1-(a\kappa)^2}}
\end{align}
for $(a\kappa)^{-1}<-1$,
\begin{align}
g(a\kappa)=\frac{\frac{\pi}{2}+\arctan\left[\frac{1}{a\kappa\sqrt{1-(a\kappa)^{-2}}}\right]}{a\kappa\sqrt{1-(a\kappa)^{-2}}}
\end{align}
for $-1<(a\kappa)^{-1}<1$, and $(a\kappa)^{-1}>1$ corresponds to the particle-dimer scattering continuum where is no three-body bound state.
These correspondences with respect to three-body bound states confirm that the counterpart of the three-boson attraction is indeed introduced to fermions by choosing $v_3$ as in Eq.~(\ref{eq:v3_appendix}).

\end{document}